\documentclass[preprint,amsmath,amssymb,aps,pra,longbibliography]{revtex4-1}

\usepackage[margin=0.75in, top=1in, bottom=0.85in]{geometry}

\usepackage[final,letterspace=-0]{microtype}
\usepackage[colorlinks=true, linkcolor=blue, urlcolor=blue, citecolor=blue, anchorcolor=blue]{hyperref}

\usepackage{mathpazo} 
\usepackage{upgreek}

\usepackage{graphicx}
\usepackage{dcolumn}
\usepackage{bm}
\usepackage{helvet}

\begin{document}

\title{Omni-resonant imaging across the visible}

\author{Layton A. Hall$^{1}$, Abbas Shiri$^{1}$, and Ayman F. Abouraddy$^{1,}$}
\email{Corresponding authors: raddy@creol.ucf.edu}
\affiliation{$^{1}$CREOL, The College of Optics \& Photonics, University of Central Florida, Orlando, Florida 32816, USA}

\begin{abstract}
Resonant field enhancement in optical cavities is provided over only narrow linewidths and for specific spatial modes. Consequently, spectrally restrictive planar Fabry-P{\'e}rot cavities have not contributed to date to white-light imaging, which necessitates a highly multimoded broadband field to satisfy the resonance condition. Here we show that introducing judicious angular-dispersion circumvents the fundamental trade-off between cavity linewidth and finesse in a Fabry-P{\'e}rot cavity by exciting a 130-nm-bandwidth achromatic resonance across the visible spectrum, which far exceeds the finesse-limited linewidth (0.5~nm), and even exceeds the free spectral range (45~nm). This omni-resonant configuration enables broadband color-imaging over a 100-nm-bandwidth in the visible with minimal spherical and chromatic aberrations. We demonstrate omni-resonant imaging using coherent and incoherent light, and spatially extended and localized fields comprising stationary and moving objects. This work paves the way to harnessing broadband resonant enhancements for spatially structured fields, as needed for example in solar windows.
\end{abstract}

\maketitle


Optical resonators, such as planar Fabry-P{\'e}rot (FP) cavities, are some of the most iconic components in all of photonics \cite{SalehBook07}, which underpin the development of lasers, and help enhance a plethora of effects that are useful in nonlinear optics \cite{Koshelev20Science}, spectroscopy and sensing \cite{Armani07Science,Gagliardi14book}, frequency standards and optical clocks \cite{Diddams01Science}, optical communications \cite{Xia07NP}, enhancing light-matter interactions in atomic physics \cite{Haroche06book}, and the detection of gravitational waves \cite{Abbott09RPP}. Two central features undergird the broad utility of optical cavities: the resonant field buildup within a cavity via optical recycling, and the narrow finesse-limited spectral response on resonance. Indeed, these two features are complimentary: the stronger the cavity-field enhancement, the narrower the resonant linewidth over which it can be harnessed, and vice versa. Although this fundamental trade-off is beneficial in ultrasensitive detection \cite{Armani07Science}, it precludes the usage of optical cavities in broadband applications such as white-light imaging. One example of the need for such a capability is transparent solar windows \cite{Davy17NE,Li22AS} in which the absorption of light in particular spectral bands is resonantly enhanced without spatially impacting the remaining spectral bands.

We pose the following question: Can an image be formed of an object with \textit{broadband light} without spectral gaps through an FP cavity having sparse, narrow-linewidth resonances? Such a configuration necessitates overcoming a dual challenge: achieving resonance with the multiplicity of transverse modes in a spatially structured field, and simultaneously maintaining the resonant response continuously over a broad spectrum. To date, this challenge has remained unresolved, and resonant enhancement remains confined to specific spatial modes at the discrete resonant wavelengths. A recent study has addressed one side of this dilemma by making use of a degenerate FP cavity design that self-images the field from one cavity mirror to the other after each roundtrip \cite{Yevgeny22Science}. Such a degenerate cavity allows for diffraction-limited imaging through the cavity, but only on resonance. Although this resonant-imaging configuration enabled coherent perfect absorption (CPA) \cite{Chong10PRL,Wan11S,Zhang12Light,Baranov17NRM} for spatially structured fields, it would be of course more useful if implemented over an extended bandwidth, ideally across the entire visible spectrum.

The challenge therefore remains to realize a \textit{broadband} resonant response in an FP cavity while maintaining the possibility of imaging through the cavity. A variety of approaches over the past few decades have tackled the long-standing challenge of severing the link between the cavity-photon lifetime and the resonant bandwidth (i.e., without sacrificing the cavity finesse). Proposals for a `white-light cavity' have suggested the use of specially prepared dispersive media to hold the roundtrip phase $\chi$ fixed over an extended bandwidth. In a conventional FP cavity, $\chi(\lambda)\!=\!\tfrac{4\pi}{\lambda}nd$, where $d$ is the cavity length, $n$ is its refractive index, and $\lambda$ is the free-space wavelength \cite{SalehBook07}. Resonances occur at discrete wavelengths $\lambda_{m}$ for which $\chi(\lambda_{m})\!=\!2\pi m$, where the resonance order $m$ is an integer. In a white-light cavity, a dispersive medium of refractive index $n(\lambda)$ is introduced into the cavity to hold the roundtrip phase $\chi(\lambda)\!=\!\tfrac{4\pi}{\lambda}n(\lambda)d\!=\!2\pi m$ over a bandwidth exceeding the bare-cavity linewidth \cite{Wicht97OC}. However, because this particular dispersive response requires an active resonant medium, the achievable spectral broadening is restricted  typically  to the MHz~scale by the resonant linewidth of the active medium \cite{Pati07PRL,Wu08PRA} (e.g., from $\sim\!17$~MHz to $\sim\!160$~MHz via  an atomic resonance \cite{Wu08PRA}, or from $\sim\!0.24$~MHz to $\sim\!1$~MHz via nonlinear Brilluoin gain \cite{Yum13JLT}). A more convenient strategy introduces into the cavity linear passive components that are dispersive (e.g., chirped Bragg mirrors or other axially structured system \cite{Yum13OC}) or diffractive (e.g., diffraction gratings \cite{Wise04CQG}) to produce a wavelength-dependent cavity length $d(\lambda)$ that maintains a wavelength-independent roundtrip phase $\chi(\lambda)\!=\!\tfrac{4\pi}{\lambda}nd(\lambda)\!=\!2\pi m$ over an extended bandwidth. However, although seemingly plausible, these approaches have been shown to deliver \textit{no} linewidth broadening \cite{Wise05PRL,Yum13OC}, which casts doubts on the entire enterprise of passive white-light cavities. Alternative proposals suggest the use of exotic phenomena such as exceptional points \cite{Soleymani22NC} and zero-index materials \cite{Suwunnarat22CP} for this task, but have not yet been implemented. Prospects for a truly `white-light' cavity thus remain on hold.

We have recently demonstrated significant spectral broadening of the finesse-limited resonant-linewidth of FP cavities using only linear passive components \cite{Shabahang17SR}. Rather than modifying the cavity itself in any way, pre-conditioning the incident field couples light to an `achromatic resonance' whose bandwidth not only far exceeds the bare-cavity linewidth, but can even exceed its free-spectral range (FSR) \cite{Shabahang17SR}. By exploiting the intrinsic angular dispersion (AD) characteristic of conventional FP resonances, `omni-resonance' has been verified using incoherent \cite{Shabahang17SR,Shabahang19OL} and coherent \cite{Shiri20OL,Shiri20APLP} light, and has produced broadband CPA in a thin-film silicon solar cell \cite{Villinger21AOM}, in graphene \cite{Jahromi21arxiv}, and even in resonant absorbing media \cite{Shabahang21JO}.

Here we demonstrate \textit{white-light omni-resonant imaging} over a 100-nm-bandwidth in the visible (from $\sim\!520$ to $\sim\!620$~nm) through a planar FP cavity having bare-cavity resonances of linewidth $\approx0.5$~nm and a free spectral range $\approx45$~nm. Crucially, the incident-field pre-conditioning necessary for converting the sparse resonant cavity spectrum into a broadband continuous omni-resonant spectrum does \textit{not} interfere with conveying the spatially structured field in an optical imaging system. We first show that introducing a judicious combination of linear and nonlinear AD into the optical field incident on an FP cavity can isolate a single cavity resonance and stretch it into a 130-nm-bandwidth achromatic resonance. The minimal chromatic and spherical aberrations that are necessary for white-light imaging are compatible with a 100-nm-bandwidth, which allows for broadband imaging through the FP cavity using coherent or incoherent light, extended or localized fields, and stationary or moving sources. This work therefore provides a new solution for harnessing broadband resonant enhancement for spatially structured fields, and has profound implications for constructing new optical cavities that sidestep conventional limitations set by the cavity photon lifetime.

\section*{Results}

\subsection*{Concept of omni-resonant imaging}

\begin{figure*}[t!]
\centering
\includegraphics[width=8.6cm]{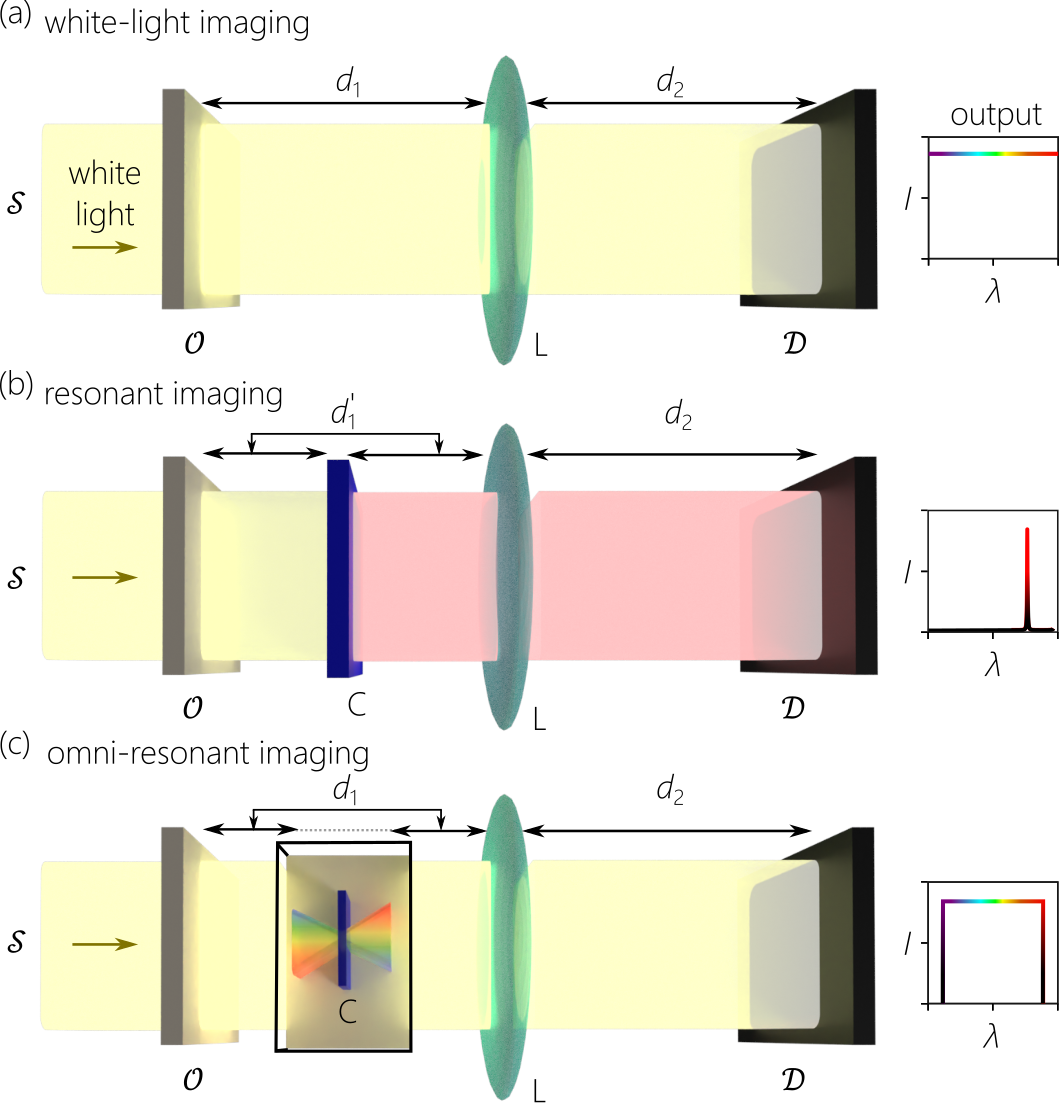}
\caption{\textbf{Conventional, resonant, and omni-resonant imaging.} (a) Conventional white-light imaging. We plot on the right the spectral transmission through the system. (b) Resonant imaging. The spectrally restricted image is formed at only the FP-cavity resonant wavelengths. The distance relevant for image-formation is $d_{1}\!=\!d_{1}'+\mathcal{F}d$, which accommodates the effective cavity length; $d$ is the cavity length, and $\mathcal{F}$ is its finesse. (c) Omni-resonant imaging. By placing the FP cavity within a system that renders it omni-resonant, broadband imaging is achieved while maintaining resonant field enhancement at all the wavelengths across the omni-resonant spectrum. The omni-resonant system is designed to be a point-to-point relay system, and does not contribute to the image-formation process. $\mathcal{S}$: Broadband source; $\mathcal{O}$: object; L: lens; C: FP cavity; $\mathcal{D}$: detector, at which an image of $\mathcal{O}$ is formed. }
\label{Fig:Concept}
\end{figure*}

To set the stage for the realization of omni-resonant imaging, we compare in Fig.~\ref{Fig:Concept} three basic configurations. In conventional `white-light imaging' [Fig.~\ref{Fig:Concept}(a)], light from a broadband source $\mathcal{S}$ scatters from an object $\mathcal{O}$, which is located at a distance $d_{1}$ from a lens L (focal length $f$), and an image is formed at distance $d_{2}$ at a detector $\mathcal{D}$, where $\tfrac{1}{d_{1}}+\tfrac{1}{d_{2}}\!=\!\tfrac{1}{f}$. The `resonant imaging' configuration in Fig.~\ref{Fig:Concept}(b), in which an FP cavity is placed in the conventional imaging system, can enhance optical effects of relevance to imaging, but most of the spectrum from $\mathcal{S}$ is rejected and image formation makes use of light only on resonance. The recently reported degenerate-cavity configuration in \cite{Yevgeny22Science} enables coupling a spatially multimoded field into an FP cavity (by placing an imaging system \textit{inside} the cavity \cite{Arnaud69AO,Tradonsky19SA}), but once again only on resonance. We aim here at image formation through the FP cavity over a broad, continuous spectrum (ideally, across the entire visible spectrum), which we refer to as `omni-resonant imaging' [Fig.~\ref{Fig:Concept}(c)] by pre-conditioning the incident field. Subsequently, the field emerging from the FP cavity is post-conditioned to produce a total point-spread function in the form of an impulse (point-by-point imaging or field relay) from the entrance of the omni-resonant system to its exit. Image formation then relies on external lenses, without the cavity reducing the useful source spectrum.

\subsection*{Exciting an achromatic resonance}

\begin{figure*}[t!]
\centering
\includegraphics[width=16cm]{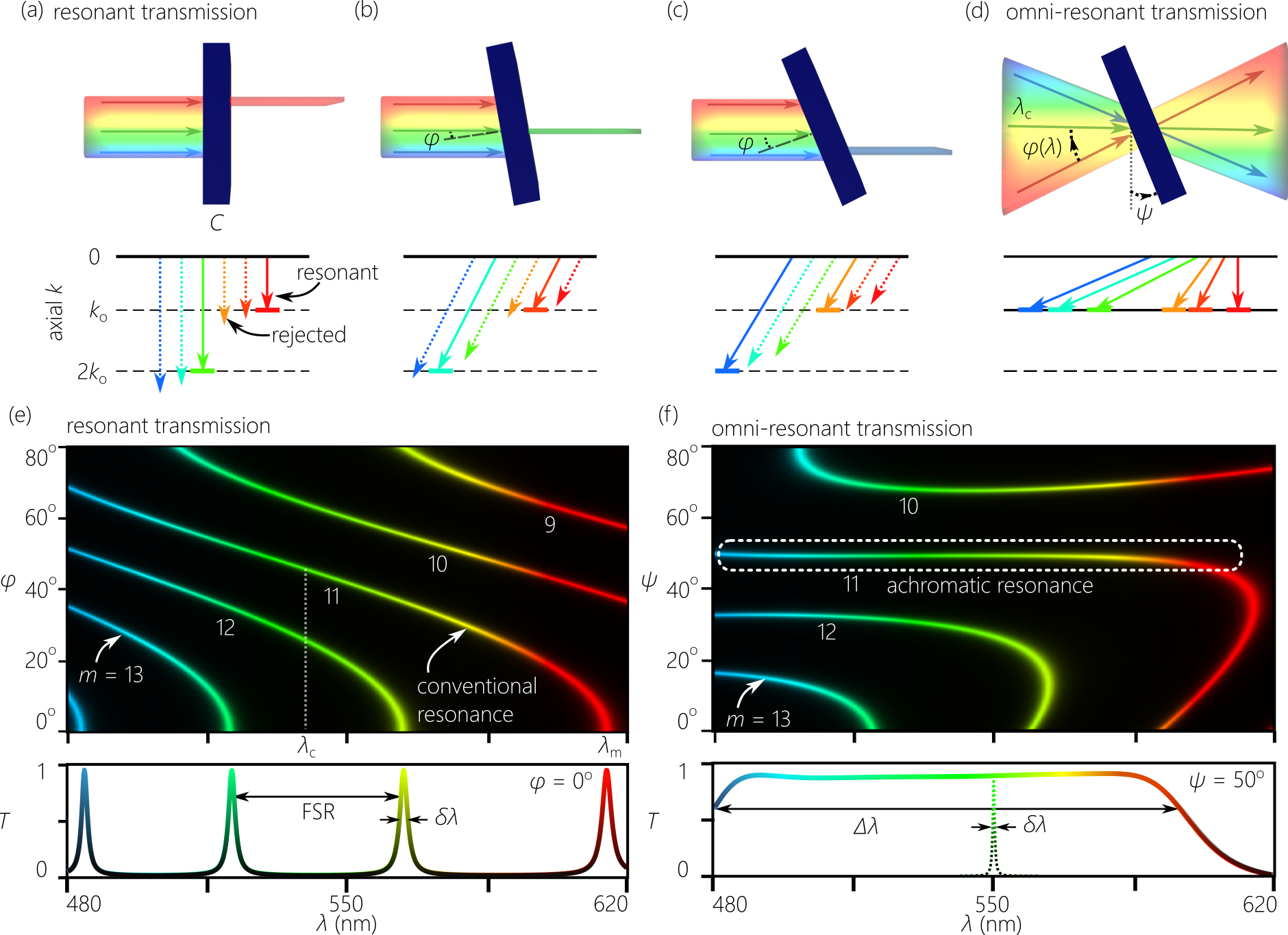}
\caption{\textbf{Resonant and omni-resonant spectral transmission.} (a-c) Resonant transmission through an FP cavity C at (a) normal and (b,c) oblique incidence, and a visualization of the axial wave number $k$ that satisfies the resonant condition. (d) The omni-resonant angular configuration incident on the FP cavity. (e) Calculated resonant transmission $T(\lambda,\varphi)$ for collimated light incident on the FP cavity; bottom panel is $T(\lambda)$ at normal incidence $\varphi\!=\!0^{\circ}$ (see text for the FP-cavity parameters). (f) Calculated omni-resonant transmission $T(\lambda,\psi)$ with the cavity tilt angle $\psi$ shown in (d), assuming the AD produced by an apsherical lens [see Fig.~\ref{Fig:Setups}(a,b) and Methods]. The achromatic resonance corresponding to $m\!=\!11$ in (e) is identified with a dashed white rectangle; bottom panel is $T(\lambda)$ at $\psi\!\approx\!50^{\circ}$.}
\label{Fig:Theory}
\end{figure*}

To elucidate the concept of omni-resonance, we first note some salient features of conventional FP resonances at oblique incidence. Consider a symmetric cavity of finesse $\mathcal{F}$ formed of two mirrors separated by a transparent layer of thickness $d$ and refractive index $n$. When collimated broadband light impinges on the cavity at an angle $\varphi$ with respect to its normal, resonances occur when $\chi(\lambda,\varphi)\!=\!2\pi m$. The wavelength associated with the $m^{\mathrm{th}}$ resonance varies widely with $\varphi$, from $\lambda_{m}\!=\!\tfrac{2nd}{m}$ at normal incidence [Fig.~\ref{Fig:Theory}(a)] to a \textit{blue-shifted} wavelength $\lambda_{\mathrm{min}}\!=\!\lambda_{m}(1-\sigma_{n})$ [Fig.~\ref{Fig:Theory}(b,c)] at glancing incidence ($\varphi\!\rightarrow\!90^{\circ}$); where we have defined for convenience the parameter $\sigma_{n}\!=\!1-\sqrt{1-\tfrac{1}{n^{2}}}$. The maximum bandwidth associated in principle with the $m^{\mathrm{th}}$-resonance is therefore $\Delta\lambda_{m}\!=\!\lambda_{m}-\lambda_{\mathrm{min}}\!=\!\sigma_{n}\lambda_{m}$, which is significantly larger than the resonant linewidth $\delta\lambda$: $\tfrac{\Delta\lambda_{m}}{\delta\lambda}\!=\!(m+1)\sigma_{n}\mathcal{F}\!\gg\!1$; and can even be larger than the FSR: $\tfrac{\Delta\lambda_{m}}{\mathrm{{FSR}}}\!=\!(m+1)\sigma_{n}$. Conventional FP resonances are therefore intrinsically characterized by AD \cite{Torres10AOP}; that is, for the $m^{\mathrm{th}}$ resonance, each wavelength $\lambda$ within $\Delta\lambda_{m}$ is associated with a particular incident angle $\varphi(\lambda)$ to satisfy the resonance condition. We plot in Fig.~\ref{Fig:Theory}(e) the spectral transmission $T(\lambda,\varphi)$ for an FP cavity with $n\!=\!1.49$, $d\!=\!2.5$~$\mu$m, and $\mathcal{F}\!=\!30$. The $m = 11$ resonance at normal incidence occurs at $\lambda_{11}\!=\!612$~nm, the FSR is $\approx\!45$~nm, and the full bandwidth associated in principle with this resonance is $\Delta\lambda_{11}\!\approx\!158$~nm (Methods).  

These characteristics of conventional FP resonances suggest a strategy for broadening the resonant spectral response associated with a fixed-order resonance and without impacting the cavity finesse. The AD profile for an FP resonance is in general nonlinear [Fig.~\ref{Fig:Theory}(e)], especially in the vicinity of $\lambda_{m}$ and $\lambda_{\mathrm{min}}$. However, the AD profile is approximately linear over a broad span between $\lambda_{m}$ and $\lambda_{\mathrm{min}}$. The condition $\tfrac{d^{2}\varphi}{d\lambda^{2}}\!=\!0$ is satisfied at a wavelength $\lambda_{\mathrm{c}}\!=\!\sqrt{\lambda_{m}\lambda_{\mathrm{min}}}$, which corresponds to an incident angle $\sin\{\varphi(\lambda_{\mathrm{c}})\}\!=\!n\sqrt{\sigma_{n}}$. The AD remains approximately linear in the vicinity of $\lambda_{\mathrm{c}}$, with a first-order AD coefficient $\beta_{m}\!=\!\frac{d\varphi}{d\lambda}\big|_{\lambda_{\mathrm{c}}}\!=\!-\frac{1}{\Delta\lambda_{m}}$. Therefore, by externally introducing linear AD into the incident optical field to match the FP-resonance [Fig.~\ref{Fig:Theory}(d)], a broad spectrum can resonate continuously with the cavity while remaining associated with the $m^{\mathrm{th}}$ resonance. This is accomplished by directing each wavelength $\lambda$ to the FP cavity at an angle $\varphi(\lambda)\!=\!\psi+\beta_{m}(\lambda-\lambda_{\mathrm{c}})$, such that the wavelength $\lambda_{\mathrm{c}}$ is incident at an angle $\psi$ onto the FP cavity, and the other wavelengths are angularly distributed linearly around $\psi$. When the cavity is tilted such that $\psi\!=\!\varphi(\lambda_{\mathrm{c}})$, which we call the omni-resonant condition, a broad spectrum centered on $\lambda_{\mathrm{c}}$ resonates continuously with the FP cavity and remains in its entirety associated with the same resonant order $m$.

We plot in Fig.~\ref{Fig:Theory}(f) the spectral transmission $T(\lambda,\psi)$ through the same FP cavity employed in Fig.~\ref{Fig:Theory}(e) after introducing an AD profile $\varphi(\lambda)\!=\!\psi+\beta_{11}(\lambda-\lambda_{\mathrm{c}})$, where $\beta_{11}\!\approx\!-0.36$~$^{\circ}$/nm and $\lambda_{\mathrm{c}}\!\approx\!530$~nm (Methods), resulting in an achromatic resonance at $\psi\!=\!\varphi(\lambda_{\mathrm{c}})\!\approx\!50^{\circ}$. Once the cavity is tilted away from this angle, the achromatic resonance disappears. The omni-resonant bandwidth associated with $m\!=\!11$ is $\Delta\lambda\!\approx\!130$~nm [Fig.~\ref{Fig:Theory}(f)]. While holding the AD coefficient $\beta_{11}$ fixed, other achromatic resonances occur while varying $\psi$ that are associated with different $m$; e.g., $m\!=\!12$ at $\psi\!\approx\!35^{\circ}$ [Fig.~\ref{Fig:Theory}(f)]. However, the bandwidths of these other achromatic resonances are sub-optimal. Finally, we note that we included in the calculation in Fig.~\ref{Fig:Theory}(f) an additional nonlinear component to $\varphi(\lambda)$, which arises in our experimental setup in Fig.~\ref{Fig:Setups}(a) and further improves the omni-resonant bandwidth over that resulting from strictly linear AD (Methods). 

\subsection*{Omni-resonance imaging implementation}

The FP cavity we make use of here consists of a pair of dielectric Bragg mirrors sandwiching a $\approx2.2$-$\upmu$m-thick SiO$_{2}$ layer. Each mirror is formed of 8 alternating bilayers of TiO$_{2}$ and SiO$_{2}$ (refractive indices of $\approx\!1.49$ and $2.28$, respectively, at $\lambda\!\sim\!540$~nm). The resulting Bragg reflectance is $\approx\!98\%$ over a 180-nm-bandwidth ($\sim\!500\!-\!680$~nm), a finesse $\mathcal{F}\!\approx\!85$, resonance linewidth $\approx0.5$~nm, free spectral range $\approx45$~nm, and normal-incidence resonant wavelengths $\lambda_{m}\approx522$, 565, and 610~nm for $m\!=\!13$, 12, and 11, respectively. The value of AD required to excite an achromatic resonance associated with $m\!=\!11$ ($\beta_{11}\!=\!-0.36$~$^{\circ}$/nm) is significantly larger than can be achieved simply by a diffraction grating \cite{SalehBook07} or a metasurface \cite{Arbabi17Optica} in the visible. Instead, we realize the requisite AD profile using the setups shown in Fig.~\ref{Fig:Setups}(a,d), which have the same overall structure. First, a diffraction grating introduces linear AD of value $\beta\!<\!\beta_{11}$, followed by an achromatic biconvex lens L$_{1}$ that collimates the spectrum at a plane denoted the `input plane'. The bandwidth intercepted by the aperture of L$_{1}$ is determined by $\beta$, and the wavelength $\lambda_{\mathrm{c}}$ coincides with the symmetry axis of the system. Following L$_{1}$, lenses are used to increase the value of $\beta$ to reach $\beta_{11}$, and the field then impinges on the FP cavity, which is tilted an angle $\psi$ with respect to the wavelength $\lambda_{\mathrm{c}}$. The optical system following the FP cavity mirrors that preceding it in order to reconstitute the spectrum into a collimated beam. An `output plane' after the FP cavity, where the white-light spectrum is spatially resolved, mirrors the location of the input plane preceding it. A measurement at this plane unveils whether conventional FP resonances (discrete, sparse spectrum) or an achromatic resonance (continuous, broadband spectrum) have been excited. Between the two gratings, the wave front is spectrally resolved in space, which necessitates careful design to minimize chromatic and spherical aberrations.

In the first setup [Fig.~\ref{Fig:Setups}(a)], the grating $G_{1}'$ produces $\beta\approx-0.05$~$^{\circ}$/nm and the bandwidth intercepted by L$_{1}$ is $\approx190$~nm. The linear AD coefficient is subsequently increased by $7\times$ using an afocal magnification system comprising  L$_{1}$ and a plano-convex aspheric condenser lens L$_{2}'$ (Methods). We plot in Fig.~\ref{Fig:Setups}(b) the calculated AD profile $\varphi(\lambda)$ produced by this setup, and compare it to the spectral trajectory of the conventional FP resonance $m\!=\!11$. Note that the predicted AD profile matches the finesse-limited FP resonance over a broad bandwidth. The AD produced after L$_{2}'$ is \textit{not} purely linear, because the local curvature along the lens surface varies with height. As shown in Fig.~\ref{Fig:Setups}(b), inset, L$_{2}'$ has a prescribed lens `sag', a height-dependent lens thickness $z(h)$, from which we calculate the local lens curvature \cite{GearyBook2002} ($h$ is the height from the optical axis). For a collimated field incident on L$_{2}'$ (prepared by L$_{1}$) we calculate the angle $\varphi(h)$ with the $z$-axis that a ray incident on the curved surface of L$_{2}'$ at height $h$ emerges from its planar surface. Because the ray at height $h$ is associated with a particular wavelength $\lambda$, we readily obtain the AD profile $\varphi(\lambda)$ (Methods). For wavelengths close to $\lambda_{\mathrm{c}}$ in the vicinity of the optical axis, $\varphi(\lambda)$ is approximately linear in $\lambda$. However, the extreme ends of the incident spectrum that reach the edges of L$_{2}'$ have an AD profile with a significant nonlinear contribution. The particular sag profile of L$_{2}'$ produces a nonlinear AD profile $\varphi(\lambda)$ away from $\lambda_{\mathrm{c}}$ that matches the spectral trajectory of the FP resonance $m\!=\!11$, thus extending the omni-resonant bandwidth beyond that afforded by linear AD.

To verify omni-resonant transmission, we direct white-light to G$_{1}'$ in absence of an object, and plot in Fig.~\ref{Fig:Setups}(c) three spatially resolved spectra recorded by a color CCD camera (Sony $\alpha6000$): (1) the spectrum at the input plane, which corresponds to an intercepted bandwidth $\approx190$~nm from the source as determined by G$_{1}'$ and the aperture of L$_{1}$; (2) the conventional resonant spectrum transmitted through the FP cavity at normal incidence in absence of AD(corresponding to resonances $m\!=\!13$, 12, and 11 at wavelengths $\approx\!522$, 565, and 610~nm, respectively); and (3) the spectrum at the output plane transmitted through the FP cavity. The conventional resonant spectrum is obtained by directing the collimated white-light source normally to the FP cavity, and then spatially resolving the transmitted spectrum via G$_{1}'$ and L$_{1}$. We observe in the output spectrum an achromatic resonance of bandwidth $\Delta\lambda\!\approx\!130$~nm associated with $m\!=\!11$. Not only does this bandwidth far exceed the resonant linewidth, but it also exceeds twice the FSR. This is the largest resonant linewidth enhancement factor reported to date using any technique.

However, calculations and measurements reveal that this system is associated with significant spherical aberrations associated with the use of the aspheric lens L$_{2}'$, especially at wavelengths far away from the central wavelength of 550~nm; see Fig.~\ref{Fig:Setups}(b) and Methods. To minimize these aberrations, we employ the setup shown in Fig.~\ref{Fig:Setups}(d) in which the grating G$_{1}$ introduces higher AD of $\beta\!\approx\!-0.09$~$^{\circ}$/nm, and we replace the aspheric lens L$_{2}'$ with a pair of achromatic lenses L$_{2}$ and L$_{3}$ that comprise an afocal imaging telescope. This system increases the AD by $4\times$ and produces AD that is closer to a linear profile; see Fig.~\ref{Fig:Setups}(e) and Methods. Calculations reveal spherical aberrations of $\sim\!0.2$~wavelengths for an entrance pupil of diameter 20~mm \cite{GearyBook2002}, thus producing an Airy radius of 36~$\upmu$m that contains within it the majority of the incident spectrum [Fig.~\ref{Fig:Setups}(e), inset]. This latter requirement is critical to achieve high-quality broadband color-imaging. The measured input, resonant, and omni-resonant spectra for the setup in Fig.~\ref{Fig:Setups}(d) are plotted in Fig.~\ref{Fig:Setups}(f). The input spectrum intercepted by L$_{1}$ is reduced to $\approx100$~nm ($\sim520-620$~nm) because of the larger AD from G$_{1}$, and an omni-resonant bandwidth $\approx100$~nm is observed at the output plane.

The full picture of the resonant and omni-resonant spectral responses are shown in Fig.~\ref{Fig:Spectra}. We first plot in Fig.~\ref{Fig:Spectra}(a,b) the measured conventional resonant transmission spectra through the FP cavity obtained using collimated white light while varying the incident angle $\varphi$, which reveals the spectral trajectories of the conventional FP resonances $m\!=\!10-13$ (Supplementary Movie~1). We then plot in Fig.~\ref{Fig:Spectra}(c,d) the measured transmission through the same cavity after introducing AD using the setup in Fig.~\ref{Fig:Setups}(a) while varying $\psi$ (Supplementary Movie~2). The optimal achromatic resonance associated with $m\!=\!11$ appears at $\psi\approx50^{\circ}$ extending over the green and red portions of the visible spectrum, while a sub-optimal achromatic resonance associated with $m\!=\!12$ appears at $\psi\approx35^{\circ}$ and dominated by the blue portion of the visible spectrum. Similar results are obtained using the setup in Fig.~\ref{Fig:Setups}(d) except for smaller bandwidths achieved for the achromatic resonances.

\subsection*{Experimental results for omni-resonant imaging}

We have confirmed above the significant resonant spectral broadening afforded by the omni-resonant configuration. Calculations for the setup in Fig.~\ref{Fig:Setups}(d) predict low imaging aberrations despite this spectral broadening, which indicates that broadband imaging can be carried out through such a system. We next verify omni-resonant imaging using objects in the form of transmissive color transparencies. The collimated white-light source is transmitted through the object, and the spatially structured field impinges on the omni-resonant FP cavity system [Fig.~\ref{Fig:Setups}(d)], and a biconvex lens (focal length $f\!=\!100$~mm) forms an image that is captured with a color camera (Sony~$\alpha6000$). Because this omni-resonant configuration corresponds to a point-to-point relay system, the axial extent from G$_{1}$ (the entrance plane) to G$_{2}$ (the exit plane) does not contribute to image formation [Fig.~\ref{Fig:Concept}(c)]. Excluding the omni-resonant system, the distance from the object plane to the lens is $d_{1}\!=\!300$~mm, and from the lens to the image plane $d_{2}\!=\!150$~mm.


We make use of three objects to test the omni-resonant imaging system, each displaying a Pegasus of dimensions $15\!\times\!15$~mm$^{2}$ colored blue, green, and red. The images captured by the color camera are displayed in Fig.~\ref{Fig:Pegasus}(a-c). In each case, we show a reference image of the object and the corresponding omni-resonant image. The reference image is acquired by placing the object \textit{after} the omni-resonant system, which simply delivers a collimated white-light field whose spectrum is filtered to a bandwidth $\Delta\lambda\approx100$~nm. The omni-resonant image, on the other hand, is acquired by placing the object \textit{before} the omni-resonant system, in which case the broadband optical field incident on G$_{1}$ is spatially structured. Nevertheless, the reference and omni-resonant images formed are identical. This indicates that the omni-resonant system does not hinder the image-formation process. Next we make use of objects in the form of a multi-colored Pegasus as shown in Fig.~\ref{Fig:Pegasus}(d,e), and once again we find excellent agreement between the reference and omni-resonant images. Finally, we make use of a black-and-white Pegasus as shown in Fig.~\ref{Fig:Pegasus}(f). The image in this last case appears yellow rather than white because the blue part of the visible spectrum is not included in the omni-resonant bandwidth [Fig.~\ref{Fig:Setups}(f)]. In the images displayed in Fig.~\ref{Fig:Pegasus}(b-f), we make use of the achromatic resonance associated with $m\!=\!11$ (at $\psi\!\approx\!50^{\circ}$). This achromatic resonance is truncated at the blue-end of the spectrum, and extends from the green ($\approx520$~nm) to the red ($\approx620$~nm). To produce the blue image in Fig.~\ref{Fig:Pegasus}(a), we made use of the sub-optimal achromatic resonance associated with $m\!=\!12$, which is excited at $\psi\!\approx\!35^{\circ}$ after adjusting the incident angle on G$_{1}$ to $15^{\circ}$ to arrange for the blue portion of the spectrum to impinge on the FP cavity. This partial achromatic resonance extends from the blue ($\sim485$~nm) to the green ($\sim530$~nm). Hence, full-color visible omni-resonant imaging has not yet been realized using a single achromatic resonance. We investigate this limit further in the Discussion.

\begin{figure*}[t!]
\centering
\includegraphics[width=8.6cm]{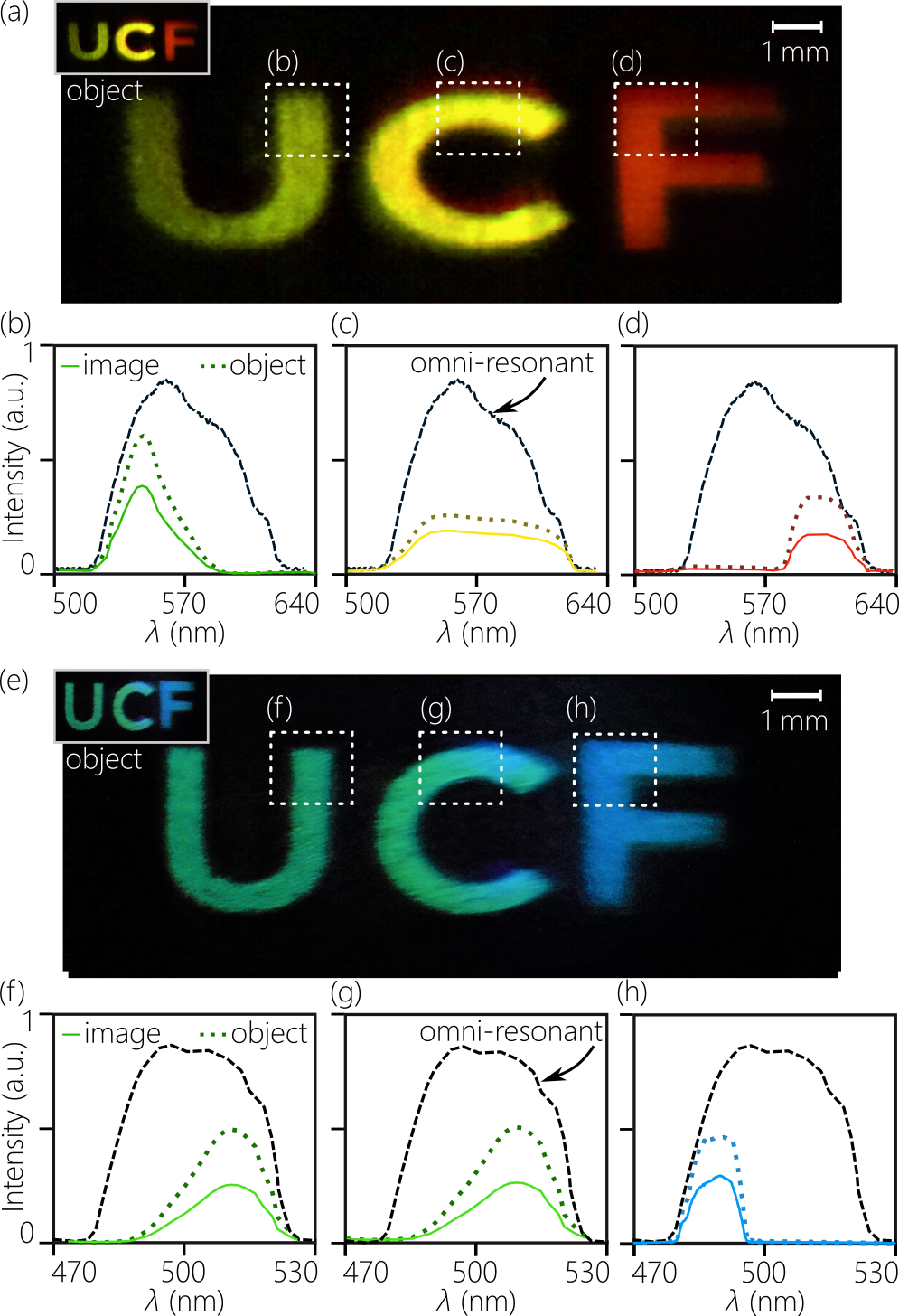}
\caption{\textbf{Measured spectra for broadband omni-resonant imaging.} (a) Omni-resonant image of a color object (the letters `UCF'). The top-left inset shows the reference image. We make use of the $m=11$ achromatic resonance at $\psi\approx50^{\circ}$. (b-d) The measured spectra within the identified white dotted boxes in (a), compared to the overall omni-resonant bandwidth, along with the measured spectrum from the corresponding regions in the reference image. (e-h) are the same as (a-d) but for the $m=12$ achromatic resonance at $\psi\!\approx\!35^{\circ}$ (omni-resonant spectrum shifted to the range $\sim\!485-530$~nm).}
\label{Fig:UCF}
\end{figure*}

To verify \textit{quantitatively} that the colors captured in the omni-resonant image correspond to those in the object, we make use of an object displaying different-colored letters `UCF' (with dimensions $10\!\times\!4$~mm$^{2}$). The omni-resonant image is shown in Fig.~\ref{Fig:UCF}(a), and the reference image is shown as the inset. We capture with a multimode fiber light from each letter corresponding to the highlighted regions in Fig.~\ref{Fig:UCF}(a), and then compare the measured spectra in the omni-resonant and reference images to the transmitted spectra through the same portions of the object when illuminated with the collimated 100-nm-bandwidth omni-resonant spectrum [Fig.~\ref{Fig:UCF}(b-d)]. The object and image spectra are almost identical, thereby confirming the spectral fidelity of the omni-resonant color-imaging system. We repeat these measurements with an object whose colors are predominantly blue [Fig.~\ref{Fig:UCF}(e)], where we made use of the $m\!=\!12$ partial achromatic resonance described above. The object and image spectra are in excellent agreement [Fig.~\ref{Fig:UCF}(f-h)].

The omni-resonant images in Fig.~\ref{Fig:Pegasus} and Fig.~\ref{Fig:UCF} made use of incoherent white light. We finally combine coherent and incoherent light in the omni-resonant imaging system [Fig.~\ref{Fig:Movies}]. The coherent light comprises a pair of laser diodes: laser~1 at a wavelength of 532~nm (Thorlabs CPS532) and laser~2 at a wavelength of 632~nm (Newport N-LHP-111). The white light and the laser beams are combined on a beam splitter. The spectrum of the combined source is measured in absence of the object at the input plane (prior to the FP cavity) and the output plane (after the FP cavity) identified in Fig.~\ref{Fig:Setups}(d), which are displayed in Fig.~\ref{Fig:Movies}(a) and Fig.~\ref{Fig:Movies}(b), respectively. The two spectra are in excellent agreement, and we can clearly discern the broadband incoherent white-light spectrum and the spectrally localized coherent lasers. Next, we insert the black-and-white Pegasus object from Fig.~\ref{Fig:Pegasus}(f) into the combined-beam path, and capture the color image as the two laser beams are scanned independently but simultaneously across the plane of the object by displacing the laser beams in the transverse plane. A Supplementary Movie captures in real time the omni-resonant color image produced by the combined source, and Fig.~\ref{Fig:Movies}(c-e) shows selected still images of the Pegasus object. The extended white-light field-of-view is stationary and illuminates the full object, whereas the green and red laser beams are spatially localized and are scanned independently across the object plane with the captured image tracking their movement in real time.


\section*{Discussion}

\noindent

We have demonstrated broadband omni-resonant imaging using color objects through a planar FP cavity. We first isolated a single conventional FP resonance (with a resonant linewidth $\delta\lambda\approx0.5$~nm and FSR $\approx45$~nm) and converted it into an achromatic resonance with extended omni-resonant bandwidth $\approx130$~nm in the visible, by introducing both linear and nonlinear AD into the incident field. This dramatic enhancement ($\approx260\times$) in the resonant bandwidth does not sacrifice the cavity finesse. However, minimizing the spherical and chromatic aberrations in the omni-resonant system restricts the achromatic resonance to a bandwidth $\approx100$~nm. Imaging results over this extended bandwidth indicate that broadband resonantly enhanced effects can be applied to arbitrary incident wave fronts, including white-light imaging (potentially across the entire visible spectrum) using either coherent or incoherent light, spatially extended or localized fields, and stationary or moving optical sources. In general, the omni-resonant imaging system is agnostic with respect to the source wavelength (as long as it lies within the omni-resonant bandwidth), location (as long as it falls within the field-of-view of the omni-resonant imaging system), or size (as long as it is within the omni-resonant imaging resolution \cite{Shiri22OL}).

Several critical questions regarding the limits of omni-resonant imaging need to be addressed. First, what is the ultimate bandwidth achievable in omni-resonant systems? Clearly, the maximum achievable omni-resonant spectrum associated entirely with a single resonance of order $m$ is $\Delta\lambda_{m}\!=\!\sigma_{n}\lambda_{m}$. This bandwidth can be quite considerable, especially for low-refractive-index materials. For example, $\Delta\lambda_{m}\!=\!272$~nm when $n\!=\!1.3$ for a normal-incidence resonant wavelength $\lambda_{m}\!=\!610$~nm. In our case here, $\Delta\lambda_{m}\!=\!158$~nm for $n\!\approx\!1.49$ and $m\!=\!11$. Moreover, we have not yet made full use of the available achromatic-resonance bandwidth $\Delta\lambda_{m}$, the full utilization of which requires sculpting the AD profile over a large bandwidth and numerical aperture. An alternative strategy is spectral recycling \cite{Hall21PRA}, in which the target omni-resonant spectrum is divided into segments, each of which is associated with a different FP resonant order. In this case, the omni-resonant spectrum can in principle extend from the ultraviolet to the mid-infrared, and is ultimately limited only by the transparency of the materials used and the bandwidth of the FP-cavity mirrors.

Second, can the entire visible spectrum ($\sim\!200$~nm) be encompassed within the omni-resonant imaging system? What is the limit for the spatial resolution that can be maintained by the omni-resonant imaging system over such a broad spectrum? We have mentioned earlier the degenerate cavity presented in Ref.~\cite{Yevgeny22Science} that provides diffraction-limited imaging through a planar FP cavity that is independent of the cavity finesse. However, this demonstration was realized only on resonance. It is an open question whether the omni-resonant strategy pursued here can benefit the massively degenerate cavity approach. This would provide broadband resonant imaging with diffraction-limited and finesse-independent spatial resolution. Moreover, resonant effects such as CPA can be combined with image formation in the omni-resonant configuration, as demonstrated in \cite{Villinger21AOM,Jahromi21arxiv,Shabahang21JO} (as long as the spectral reflectance of the cavity mirror \cite{Villinger15OL,Pye17OL} accommodates the spectral change in absorption that occurs over extended bandwidths).

Finally, because the omni-resonant effect can be implemented with both coherent and incoherent light, one may envision a scenario where broadband resonantly enhanced nonlinear effects may be realized when using ultrashort pulses. One avenue that can benefit from such an approach is reflective nonlinear limiters \cite{Makri14PRA,Vella16PRAppl}. Moreover, the various logic operations that can be performed in the linear regime \cite{Fang15Light} may now make use of broadband sources.

In our experiments we made use of bulk optical components (e.g., gratings and lenses), which may limit the practical applications of omni-resonant imaging. In principle, such components may be replaced with a fewer number of devices based on flat-optics technologies; such as diffractive optics \cite{Banerji19Optica,Meem20Optica}, free-form optics \cite{Nikolov21SA}, or metasurfaces \cite{Yu14NM,Dorrah22Science}. It remains an open question whether a \textit{single-surface} element or a few surfaces are required to introduce the requisite AD to achieve omni-resonance across the entire visible, and what is the minimum system volume needed for this \cite{Miller23Science}. A further challenge lies in designing optics for aberration-free omni-resonant imaging across the entire visible. Whereas an aspherical lens provides the appropriate AD across a broad bandwidth, it nevertheless introduces imaging aberrations at the extremities of the spectrum. Although such aberrations can be corrected in principle at any wavelength via an appropriate phase plate, they cannot be fully corrected over the entire visible via a single phase plate. An important question that emerges here is therefore whether the flexibility afforded by metasurfaces can enable the accomplishment of both tasks (satisfying the omni-resonant condition while simultaneously correcting for spherical and chromatic aberrations) using a single optical element. Addressing these challenges would enable for nonlinear optical effects in large-area, thin-film devices, and transparent solar windows, among a host of potential opportunities.


\bibliography{diffraction}

\begin{thebibliography}{49}%
\makeatletter
\providecommand \@ifxundefined [1]{%
 \@ifx{#1\undefined}
}%
\providecommand \@ifnum [1]{%
 \ifnum #1\expandafter \@firstoftwo
 \else \expandafter \@secondoftwo
 \fi
}%
\providecommand \@ifx [1]{%
 \ifx #1\expandafter \@firstoftwo
 \else \expandafter \@secondoftwo
 \fi
}%
\providecommand \natexlab [1]{#1}%
\providecommand \enquote  [1]{``#1''}%
\providecommand \bibnamefont  [1]{#1}%
\providecommand \bibfnamefont [1]{#1}%
\providecommand \citenamefont [1]{#1}%
\providecommand \href@noop [0]{\@secondoftwo}%
\providecommand \href [0]{\begingroup \@sanitize@url \@href}%
\providecommand \@href[1]{\@@startlink{#1}\@@href}%
\providecommand \@@href[1]{\endgroup#1\@@endlink}%
\providecommand \@sanitize@url [0]{\catcode `\\12\catcode `\$12\catcode
  `\&12\catcode `\#12\catcode `\^12\catcode `\_12\catcode `\%12\relax}%
\providecommand \@@startlink[1]{}%
\providecommand \@@endlink[0]{}%
\providecommand \url  [0]{\begingroup\@sanitize@url \@url }%
\providecommand \@url [1]{\endgroup\@href {#1}{\urlprefix }}%
\providecommand \urlprefix  [0]{URL }%
\providecommand \Eprint [0]{\href }%
\providecommand \doibase [0]{http://dx.doi.org/}%
\providecommand \selectlanguage [0]{\@gobble}%
\providecommand \bibinfo  [0]{\@secondoftwo}%
\providecommand \bibfield  [0]{\@secondoftwo}%
\providecommand \translation [1]{[#1]}%
\providecommand \BibitemOpen [0]{}%
\providecommand \bibitemStop [0]{}%
\providecommand \bibitemNoStop [0]{.\EOS\space}%
\providecommand \EOS [0]{\spacefactor3000\relax}%
\providecommand \BibitemShut  [1]{\csname bibitem#1\endcsname}%
\let\auto@bib@innerbib\@empty
\bibitem [{\citenamefont {Saleh}\ and\ \citenamefont
  {Teich}(2007)}]{SalehBook07}%
  \BibitemOpen
  \bibfield  {author} {\bibinfo {author} {\bibfnamefont {B.~E.~A.}\
  \bibnamefont {Saleh}}\ and\ \bibinfo {author} {\bibfnamefont {M.~C.}\
  \bibnamefont {Teich}},\ }\href@noop {} {\emph {\bibinfo {title} {Principles
  of Photonics}}}\ (\bibinfo  {publisher} {Wiley},\ \bibinfo {year}
  {2007})\BibitemShut {NoStop}%
\bibitem [{\citenamefont {Koshelev}\ \emph {et~al.}(2020)\citenamefont
  {Koshelev}, \citenamefont {Kruk}, \citenamefont {Melik-Gaykazyan},
  \citenamefont {Choi}, \citenamefont {Bogdanov}, \citenamefont {Park},\ and\
  \citenamefont {Kivshar}}]{Koshelev20Science}%
  \BibitemOpen
  \bibfield  {author} {\bibinfo {author} {\bibfnamefont {K.}~\bibnamefont
  {Koshelev}}, \bibinfo {author} {\bibfnamefont {S.}~\bibnamefont {Kruk}},
  \bibinfo {author} {\bibfnamefont {E.}~\bibnamefont {Melik-Gaykazyan}},
  \bibinfo {author} {\bibfnamefont {J.-H.}\ \bibnamefont {Choi}}, \bibinfo
  {author} {\bibfnamefont {A.}~\bibnamefont {Bogdanov}}, \bibinfo {author}
  {\bibfnamefont {H.-G.}\ \bibnamefont {Park}}, \ and\ \bibinfo {author}
  {\bibfnamefont {Y.}~\bibnamefont {Kivshar}},\ }\bibfield  {title} {\enquote
  {\bibinfo {title} {Subwavelength dielectric resonators for nonlinear
  nanophotonics},}\ }\href@noop {} {\bibfield  {journal} {\bibinfo  {journal}
  {Science}\ }\textbf {\bibinfo {volume} {367}},\ \bibinfo {pages} {288--292}
  (\bibinfo {year} {2020})}\BibitemShut {NoStop}%
\bibitem [{\citenamefont {Armani}\ \emph {et~al.}(2007)\citenamefont {Armani},
  \citenamefont {Kulkarni}, \citenamefont {Fraser}, \citenamefont {Flagan},\
  and\ \citenamefont {Vahala}}]{Armani07Science}%
  \BibitemOpen
  \bibfield  {author} {\bibinfo {author} {\bibfnamefont {A.~M.}\ \bibnamefont
  {Armani}}, \bibinfo {author} {\bibfnamefont {R.~P.}\ \bibnamefont
  {Kulkarni}}, \bibinfo {author} {\bibfnamefont {S.~E.}\ \bibnamefont
  {Fraser}}, \bibinfo {author} {\bibfnamefont {R.~C.}\ \bibnamefont {Flagan}},
  \ and\ \bibinfo {author} {\bibfnamefont {K.~J.}\ \bibnamefont {Vahala}},\
  }\bibfield  {title} {\enquote {\bibinfo {title} {Label-free, single-molecule
  detection with optical microcavities},}\ }\href@noop {} {\bibfield  {journal}
  {\bibinfo  {journal} {Science}\ }\textbf {\bibinfo {volume} {317}},\ \bibinfo
  {pages} {783--787} (\bibinfo {year} {2007})}\BibitemShut {NoStop}%
\bibitem [{\citenamefont {Gagliardi}\ and\ \citenamefont
  {Loock}(2014)}]{Gagliardi14book}%
  \BibitemOpen
  \bibinfo {editor} {\bibfnamefont {G.}~\bibnamefont {Gagliardi}}\ and\
  \bibinfo {editor} {\bibfnamefont {H.-P.}\ \bibnamefont {Loock}},\ eds.,\
  \href@noop {} {\emph {\bibinfo {title} {Cavity-Enhanced Spectroscopy and
  Sensing}}}\ (\bibinfo  {publisher} {Springer},\ \bibinfo {year}
  {2014})\BibitemShut {NoStop}%
\bibitem [{\citenamefont {Diddams}\ \emph {et~al.}(2001)\citenamefont
  {Diddams}, \citenamefont {Udem}, \citenamefont {Bergquist}, \citenamefont
  {Curtis}, \citenamefont {Drullinger}, \citenamefont {Hollberg}, \citenamefont
  {Itano}, \citenamefont {Lee}, \citenamefont {Oates}, \citenamefont {Vogel},\
  and\ \citenamefont {Wineland}}]{Diddams01Science}%
  \BibitemOpen
  \bibfield  {author} {\bibinfo {author} {\bibfnamefont {S.~A.}\ \bibnamefont
  {Diddams}}, \bibinfo {author} {\bibfnamefont {T.}~\bibnamefont {Udem}},
  \bibinfo {author} {\bibfnamefont {J.~C.}\ \bibnamefont {Bergquist}}, \bibinfo
  {author} {\bibfnamefont {E.~A.}\ \bibnamefont {Curtis}}, \bibinfo {author}
  {\bibfnamefont {R.~E.}\ \bibnamefont {Drullinger}}, \bibinfo {author}
  {\bibfnamefont {L.}~\bibnamefont {Hollberg}}, \bibinfo {author}
  {\bibfnamefont {W.~M.}\ \bibnamefont {Itano}}, \bibinfo {author}
  {\bibfnamefont {W.~D.}\ \bibnamefont {Lee}}, \bibinfo {author} {\bibfnamefont
  {C.~W.}\ \bibnamefont {Oates}}, \bibinfo {author} {\bibfnamefont {K.~R.}\
  \bibnamefont {Vogel}}, \ and\ \bibinfo {author} {\bibfnamefont {D.~J.}\
  \bibnamefont {Wineland}},\ }\bibfield  {title} {\enquote {\bibinfo {title}
  {An optical clock based on a single trapped $^{199}$hg$^{+}$ ion},}\
  }\href@noop {} {\bibfield  {journal} {\bibinfo  {journal} {Science}\ }\textbf
  {\bibinfo {volume} {293}},\ \bibinfo {pages} {825--293} (\bibinfo {year}
  {2001})}\BibitemShut {NoStop}%
\bibitem [{\citenamefont {Xia}\ \emph {et~al.}(2007)\citenamefont {Xia},
  \citenamefont {Sekaric},\ and\ \citenamefont {Vlasov}}]{Xia07NP}%
  \BibitemOpen
  \bibfield  {author} {\bibinfo {author} {\bibfnamefont {F.}~\bibnamefont
  {Xia}}, \bibinfo {author} {\bibfnamefont {L.}~\bibnamefont {Sekaric}}, \ and\
  \bibinfo {author} {\bibfnamefont {Y.}~\bibnamefont {Vlasov}},\ }\bibfield
  {title} {\enquote {\bibinfo {title} {Ultracompact optical buffers on a
  silicon chip},}\ }\href@noop {} {\bibfield  {journal} {\bibinfo  {journal}
  {Nat. Photon.}\ }\textbf {\bibinfo {volume} {1}},\ \bibinfo {pages} {65--71}
  (\bibinfo {year} {2007})}\BibitemShut {NoStop}%
\bibitem [{\citenamefont {Haroche}\ and\ \citenamefont
  {Raimond}(2006)}]{Haroche06book}%
  \BibitemOpen
  \bibfield  {author} {\bibinfo {author} {\bibfnamefont {S.}~\bibnamefont
  {Haroche}}\ and\ \bibinfo {author} {\bibfnamefont {J.-M.}\ \bibnamefont
  {Raimond}},\ }\href@noop {} {\emph {\bibinfo {title} {Exploring the Quantum:
  Atoms, Cavities, and Photons}}}\ (\bibinfo  {publisher} {Oxford Univ.
  Press},\ \bibinfo {year} {2006})\BibitemShut {NoStop}%
\bibitem [{\citenamefont {Abbott{, $et~al$.}}(2009)}]{Abbott09RPP}%
  \BibitemOpen
  \bibfield  {author} {\bibinfo {author} {\bibfnamefont {B.~P.}\ \bibnamefont
  {Abbott{, $et~al$.}}},\ }\bibfield  {title} {\enquote {\bibinfo {title}
  {{LIGO:} {T}he laser interferometer gravitational-wave observatory},}\
  }\href@noop {} {\bibfield  {journal} {\bibinfo  {journal} {Rep. Prog. Phys.}\
  }\textbf {\bibinfo {volume} {72}},\ \bibinfo {pages} {076901} (\bibinfo
  {year} {2009})}\BibitemShut {NoStop}%
\bibitem [{\citenamefont {Davy}\ \emph {et~al.}(2017)\citenamefont {Davy},
  \citenamefont {Sezen-Edmonds}, \citenamefont {Gao}, \citenamefont {Lin},
  \citenamefont {Liu}, \citenamefont {Yao}, \citenamefont {Kahn},\ and\
  \citenamefont {Loo}}]{Davy17NE}%
  \BibitemOpen
  \bibfield  {author} {\bibinfo {author} {\bibfnamefont {N.~C.}\ \bibnamefont
  {Davy}}, \bibinfo {author} {\bibfnamefont {M.}~\bibnamefont {Sezen-Edmonds}},
  \bibinfo {author} {\bibfnamefont {J.}~\bibnamefont {Gao}}, \bibinfo {author}
  {\bibfnamefont {X.}~\bibnamefont {Lin}}, \bibinfo {author} {\bibfnamefont
  {A.}~\bibnamefont {Liu}}, \bibinfo {author} {\bibfnamefont {N.}~\bibnamefont
  {Yao}}, \bibinfo {author} {\bibfnamefont {A.}~\bibnamefont {Kahn}}, \ and\
  \bibinfo {author} {\bibfnamefont {Y.-L.}\ \bibnamefont {Loo}},\ }\bibfield
  {title} {\enquote {\bibinfo {title} {Pairing of near-ultraviolet solar cells
  with electrochromic windows for smart management of the solar spectrum},}\
  }\href@noop {} {\bibfield  {journal} {\bibinfo  {journal} {Nat. Energy}\
  }\textbf {\bibinfo {volume} {2}},\ \bibinfo {pages} {17104} (\bibinfo {year}
  {2017})}\BibitemShut {NoStop}%
\bibitem [{\citenamefont {Li}\ \emph {et~al.}(2022)\citenamefont {Li},
  \citenamefont {Lin}, \citenamefont {Huang}, \citenamefont {Hur},
  \citenamefont {Huang},\ and\ \citenamefont {Yao}}]{Li22AS}%
  \BibitemOpen
  \bibfield  {author} {\bibinfo {author} {\bibfnamefont {W.}~\bibnamefont
  {Li}}, \bibinfo {author} {\bibfnamefont {C.}~\bibnamefont {Lin}}, \bibinfo
  {author} {\bibfnamefont {G.}~\bibnamefont {Huang}}, \bibinfo {author}
  {\bibfnamefont {J.}~\bibnamefont {Hur}}, \bibinfo {author} {\bibfnamefont
  {B.}~\bibnamefont {Huang}}, \ and\ \bibinfo {author} {\bibfnamefont
  {S.}~\bibnamefont {Yao}},\ }\bibfield  {title} {\enquote {\bibinfo {title}
  {Selective solar harvesting windows for full-spectrum utilization},}\
  }\href@noop {} {\bibfield  {journal} {\bibinfo  {journal} {Adv. Sci.}\
  }\textbf {\bibinfo {volume} {9}},\ \bibinfo {pages} {2201738} (\bibinfo
  {year} {2022})}\BibitemShut {NoStop}%
\bibitem [{\citenamefont {Yevgeny}\ \emph {et~al.}(2022)\citenamefont
  {Yevgeny}, \citenamefont {Gil}, \citenamefont {Helmut}, \citenamefont
  {Kevin}, \citenamefont {Stefan},\ and\ \citenamefont
  {Ori}}]{Yevgeny22Science}%
  \BibitemOpen
  \bibfield  {author} {\bibinfo {author} {\bibfnamefont {S.}~\bibnamefont
  {Yevgeny}}, \bibinfo {author} {\bibfnamefont {W.}~\bibnamefont {Gil}},
  \bibinfo {author} {\bibfnamefont {H.}~\bibnamefont {Helmut}}, \bibinfo
  {author} {\bibfnamefont {P.}~\bibnamefont {Kevin}}, \bibinfo {author}
  {\bibfnamefont {R.}~\bibnamefont {Stefan}}, \ and\ \bibinfo {author}
  {\bibfnamefont {K.}~\bibnamefont {Ori}},\ }\bibfield  {title} {\enquote
  {\bibinfo {title} {Massively degenerate coherent perfect absorber for
  arbitrary wavefronts},}\ }\href@noop {} {\bibfield  {journal} {\bibinfo
  {journal} {Science}\ }\textbf {\bibinfo {volume} {377}},\ \bibinfo {pages}
  {995--998} (\bibinfo {year} {2022})}\BibitemShut {NoStop}%
\bibitem [{\citenamefont {Chong}\ \emph {et~al.}(2010)\citenamefont {Chong},
  \citenamefont {Ge}, \citenamefont {Cao},\ and\ \citenamefont
  {Stone}}]{Chong10PRL}%
  \BibitemOpen
  \bibfield  {author} {\bibinfo {author} {\bibfnamefont {Y.~D.}\ \bibnamefont
  {Chong}}, \bibinfo {author} {\bibfnamefont {L.}~\bibnamefont {Ge}}, \bibinfo
  {author} {\bibfnamefont {H.}~\bibnamefont {Cao}}, \ and\ \bibinfo {author}
  {\bibfnamefont {A.~D.}\ \bibnamefont {Stone}},\ }\bibfield  {title} {\enquote
  {\bibinfo {title} {Coherent perfect absorbers: Time-reversed lasers},}\
  }\href@noop {} {\bibfield  {journal} {\bibinfo  {journal} {Phys. Rev. Lett.}\
  }\textbf {\bibinfo {volume} {105}},\ \bibinfo {pages} {053901} (\bibinfo
  {year} {2010})}\BibitemShut {NoStop}%
\bibitem [{\citenamefont {Wan}\ \emph {et~al.}(2011)\citenamefont {Wan},
  \citenamefont {Chong}, \citenamefont {Ge}, \citenamefont {Noh}, \citenamefont
  {Stone},\ and\ \citenamefont {Cao}}]{Wan11S}%
  \BibitemOpen
  \bibfield  {author} {\bibinfo {author} {\bibfnamefont {W.}~\bibnamefont
  {Wan}}, \bibinfo {author} {\bibfnamefont {Y.}~\bibnamefont {Chong}}, \bibinfo
  {author} {\bibfnamefont {L.}~\bibnamefont {Ge}}, \bibinfo {author}
  {\bibfnamefont {H.}~\bibnamefont {Noh}}, \bibinfo {author} {\bibfnamefont
  {A.~D.}\ \bibnamefont {Stone}}, \ and\ \bibinfo {author} {\bibfnamefont
  {H.}~\bibnamefont {Cao}},\ }\bibfield  {title} {\enquote {\bibinfo {title}
  {Time-reversed lasing and interferometric control of absorption},}\
  }\href@noop {} {\bibfield  {journal} {\bibinfo  {journal} {Science}\ }\textbf
  {\bibinfo {volume} {331}},\ \bibinfo {pages} {889--892} (\bibinfo {year}
  {2011})}\BibitemShut {NoStop}%
\bibitem [{\citenamefont {Zhang}\ \emph {et~al.}(2012)\citenamefont {Zhang},
  \citenamefont {MacDonald},\ and\ \citenamefont {Zheludev}}]{Zhang12Light}%
  \BibitemOpen
  \bibfield  {author} {\bibinfo {author} {\bibfnamefont {J.}~\bibnamefont
  {Zhang}}, \bibinfo {author} {\bibfnamefont {K.~F.}\ \bibnamefont
  {MacDonald}}, \ and\ \bibinfo {author} {\bibfnamefont {N.~I.}\ \bibnamefont
  {Zheludev}},\ }\bibfield  {title} {\enquote {\bibinfo {title} {Controlling
  light-with-light without nonlinearity},}\ }\href@noop {} {\bibfield
  {journal} {\bibinfo  {journal} {Light Sci. \& Appl.}\ }\textbf {\bibinfo
  {volume} {1}},\ \bibinfo {pages} {e18} (\bibinfo {year} {2012})}\BibitemShut
  {NoStop}%
\bibitem [{\citenamefont {Baranov}\ \emph {et~al.}(2017)\citenamefont
  {Baranov}, \citenamefont {Krasnok}, \citenamefont {Shegai}, \citenamefont
  {Al{\'u}},\ and\ \citenamefont {Chong}}]{Baranov17NRM}%
  \BibitemOpen
  \bibfield  {author} {\bibinfo {author} {\bibfnamefont {D.~G.}\ \bibnamefont
  {Baranov}}, \bibinfo {author} {\bibfnamefont {A.}~\bibnamefont {Krasnok}},
  \bibinfo {author} {\bibfnamefont {T.}~\bibnamefont {Shegai}}, \bibinfo
  {author} {\bibfnamefont {A.}~\bibnamefont {Al{\'u}}}, \ and\ \bibinfo
  {author} {\bibfnamefont {Y.}~\bibnamefont {Chong}},\ }\bibfield  {title}
  {\enquote {\bibinfo {title} {Coherent perfect absorbers: linear control of
  light with light},}\ }\href@noop {} {\bibfield  {journal} {\bibinfo
  {journal} {Nat. Rev. Mater.}\ }\textbf {\bibinfo {volume} {2}},\ \bibinfo
  {pages} {17064} (\bibinfo {year} {2017})}\BibitemShut {NoStop}%
\bibitem [{\citenamefont {Wicht}\ \emph {et~al.}(1997)\citenamefont {Wicht},
  \citenamefont {Danzmann}, \citenamefont {Fleischhauer}, \citenamefont
  {Scully}, \citenamefont {M{\"u}ller},\ and\ \citenamefont
  {Rinkleff}}]{Wicht97OC}%
  \BibitemOpen
  \bibfield  {author} {\bibinfo {author} {\bibfnamefont {A.}~\bibnamefont
  {Wicht}}, \bibinfo {author} {\bibfnamefont {K.}~\bibnamefont {Danzmann}},
  \bibinfo {author} {\bibfnamefont {M.}~\bibnamefont {Fleischhauer}}, \bibinfo
  {author} {\bibfnamefont {M.}~\bibnamefont {Scully}}, \bibinfo {author}
  {\bibfnamefont {G.}~\bibnamefont {M{\"u}ller}}, \ and\ \bibinfo {author}
  {\bibfnamefont {R.-H.}\ \bibnamefont {Rinkleff}},\ }\bibfield  {title}
  {\enquote {\bibinfo {title} {White-light cavities, atomic phase coherence,
  and gravitational wave detectors},}\ }\href@noop {} {\bibfield  {journal}
  {\bibinfo  {journal} {Opt. Commun.}\ }\textbf {\bibinfo {volume} {134}},\
  \bibinfo {pages} {431--439} (\bibinfo {year} {1997})}\BibitemShut {NoStop}%
\bibitem [{\citenamefont {Pati}\ \emph {et~al.}(2007)\citenamefont {Pati},
  \citenamefont {Salit}, \citenamefont {Salit},\ and\ \citenamefont
  {Shahriar}}]{Pati07PRL}%
  \BibitemOpen
  \bibfield  {author} {\bibinfo {author} {\bibfnamefont {G.~S.}\ \bibnamefont
  {Pati}}, \bibinfo {author} {\bibfnamefont {M.}~\bibnamefont {Salit}},
  \bibinfo {author} {\bibfnamefont {K.}~\bibnamefont {Salit}}, \ and\ \bibinfo
  {author} {\bibfnamefont {M.~S.}\ \bibnamefont {Shahriar}},\ }\bibfield
  {title} {\enquote {\bibinfo {title} {Demonstration of a tunable-bandwidth
  white-light interferometer using anomalous dispersion in atomic vapor},}\
  }\href@noop {} {\bibfield  {journal} {\bibinfo  {journal} {Phys. Rev. Lett.}\
  }\textbf {\bibinfo {volume} {99}},\ \bibinfo {pages} {133601} (\bibinfo
  {year} {2007})}\BibitemShut {NoStop}%
\bibitem [{\citenamefont {Wu}\ and\ \citenamefont {Xiao}(2008)}]{Wu08PRA}%
  \BibitemOpen
  \bibfield  {author} {\bibinfo {author} {\bibfnamefont {H.}~\bibnamefont
  {Wu}}\ and\ \bibinfo {author} {\bibfnamefont {M.}~\bibnamefont {Xiao}},\
  }\bibfield  {title} {\enquote {\bibinfo {title} {White-light cavity with
  competing linear and nonlinear dispersions},}\ }\href@noop {} {\bibfield
  {journal} {\bibinfo  {journal} {Phys. Rev. A}\ }\textbf {\bibinfo {volume}
  {77}},\ \bibinfo {pages} {031801(R)} (\bibinfo {year} {2008})}\BibitemShut
  {NoStop}%
\bibitem [{\citenamefont {Yum}\ \emph {et~al.}(2013{\natexlab{a}})\citenamefont
  {Yum}, \citenamefont {Sheuer}, \citenamefont {Salit}, \citenamefont
  {Hemmer},\ and\ \citenamefont {Shahriar}}]{Yum13JLT}%
  \BibitemOpen
  \bibfield  {author} {\bibinfo {author} {\bibfnamefont {H.~N.}\ \bibnamefont
  {Yum}}, \bibinfo {author} {\bibfnamefont {J.}~\bibnamefont {Sheuer}},
  \bibinfo {author} {\bibfnamefont {M.}~\bibnamefont {Salit}}, \bibinfo
  {author} {\bibfnamefont {P.~R.}\ \bibnamefont {Hemmer}}, \ and\ \bibinfo
  {author} {\bibfnamefont {M.~S.}\ \bibnamefont {Shahriar}},\ }\bibfield
  {title} {\enquote {\bibinfo {title} {Demonstration of white light cavity
  effect using stimulated {B}rillouin scattering in a fiber loop},}\
  }\href@noop {} {\bibfield  {journal} {\bibinfo  {journal} {J. Lightwave
  Technol.}\ }\textbf {\bibinfo {volume} {32}},\ \bibinfo {pages} {3865--3872}
  (\bibinfo {year} {2013}{\natexlab{a}})}\BibitemShut {NoStop}%
\bibitem [{\citenamefont {Yum}\ \emph {et~al.}(2013{\natexlab{b}})\citenamefont
  {Yum}, \citenamefont {Liu}, \citenamefont {Hemmer}, \citenamefont {Scheuer},\
  and\ \citenamefont {Shahriar}}]{Yum13OC}%
  \BibitemOpen
  \bibfield  {author} {\bibinfo {author} {\bibfnamefont {H.~N.}\ \bibnamefont
  {Yum}}, \bibinfo {author} {\bibfnamefont {X.}~\bibnamefont {Liu}}, \bibinfo
  {author} {\bibfnamefont {P.~R.}\ \bibnamefont {Hemmer}}, \bibinfo {author}
  {\bibfnamefont {J.}~\bibnamefont {Scheuer}}, \ and\ \bibinfo {author}
  {\bibfnamefont {M.~S.}\ \bibnamefont {Shahriar}},\ }\bibfield  {title}
  {\enquote {\bibinfo {title} {The fundamental limitations on the practical
  realizations of white light cavities},}\ }\href@noop {} {\bibfield  {journal}
  {\bibinfo  {journal} {Opt. Commun.}\ }\textbf {\bibinfo {volume} {305}},\
  \bibinfo {pages} {260--266} (\bibinfo {year}
  {2013}{\natexlab{b}})}\BibitemShut {NoStop}%
\bibitem [{\citenamefont {Wise}\ \emph {et~al.}(2004)\citenamefont {Wise},
  \citenamefont {Mueller}, \citenamefont {Reitze}, \citenamefont {Tanner},\
  and\ \citenamefont {Whiting}}]{Wise04CQG}%
  \BibitemOpen
  \bibfield  {author} {\bibinfo {author} {\bibfnamefont {S.}~\bibnamefont
  {Wise}}, \bibinfo {author} {\bibfnamefont {G.}~\bibnamefont {Mueller}},
  \bibinfo {author} {\bibfnamefont {D.}~\bibnamefont {Reitze}}, \bibinfo
  {author} {\bibfnamefont {D.~B.}\ \bibnamefont {Tanner}}, \ and\ \bibinfo
  {author} {\bibfnamefont {B.~F.}\ \bibnamefont {Whiting}},\ }\bibfield
  {title} {\enquote {\bibinfo {title} {Linewidth-broadened {F}abry-{P}erot
  cavities within future gravitational wave detectors},}\ }\href@noop {}
  {\bibfield  {journal} {\bibinfo  {journal} {Class. Quant. Grav.}\ }\textbf
  {\bibinfo {volume} {21}},\ \bibinfo {pages} {S1031} (\bibinfo {year}
  {2004})}\BibitemShut {NoStop}%
\bibitem [{\citenamefont {Wise}\ \emph {et~al.}(2005)\citenamefont {Wise},
  \citenamefont {Quetschke}, \citenamefont {Deshpande}, \citenamefont
  {Mueller}, \citenamefont {Reitze}, \citenamefont {Tanner}, \citenamefont
  {Whiting}, \citenamefont {Chen}, \citenamefont {T{\"u}nnermann},
  \citenamefont {Kley},\ and\ \citenamefont {Clausnitzer}}]{Wise05PRL}%
  \BibitemOpen
  \bibfield  {author} {\bibinfo {author} {\bibfnamefont {S.}~\bibnamefont
  {Wise}}, \bibinfo {author} {\bibfnamefont {V.}~\bibnamefont {Quetschke}},
  \bibinfo {author} {\bibfnamefont {A.~J.}\ \bibnamefont {Deshpande}}, \bibinfo
  {author} {\bibfnamefont {G.}~\bibnamefont {Mueller}}, \bibinfo {author}
  {\bibfnamefont {D.~H.}\ \bibnamefont {Reitze}}, \bibinfo {author}
  {\bibfnamefont {D.~B.}\ \bibnamefont {Tanner}}, \bibinfo {author}
  {\bibfnamefont {B.~F.}\ \bibnamefont {Whiting}}, \bibinfo {author}
  {\bibfnamefont {Y.}~\bibnamefont {Chen}}, \bibinfo {author} {\bibfnamefont
  {A.}~\bibnamefont {T{\"u}nnermann}}, \bibinfo {author} {\bibfnamefont
  {E.}~\bibnamefont {Kley}}, \ and\ \bibinfo {author} {\bibfnamefont
  {T.}~\bibnamefont {Clausnitzer}},\ }\bibfield  {title} {\enquote {\bibinfo
  {title} {Phase effects in the diffraction of light: {B}eyond the grating
  equation},}\ }\href@noop {} {\bibfield  {journal} {\bibinfo  {journal} {Phys.
  Rev. Lett.}\ }\textbf {\bibinfo {volume} {95}},\ \bibinfo {pages} {013901}
  (\bibinfo {year} {2005})}\BibitemShut {NoStop}%
\bibitem [{\citenamefont {Soleymani}\ \emph {et~al.}(2022)\citenamefont
  {Soleymani}, \citenamefont {Zhong}, \citenamefont {Mokim}, \citenamefont
  {Rotter}, \citenamefont {El-Ganainy},\ and\ \citenamefont
  {{\"O}zdemir}}]{Soleymani22NC}%
  \BibitemOpen
  \bibfield  {author} {\bibinfo {author} {\bibfnamefont {S.}~\bibnamefont
  {Soleymani}}, \bibinfo {author} {\bibfnamefont {Q.}~\bibnamefont {Zhong}},
  \bibinfo {author} {\bibfnamefont {M.}~\bibnamefont {Mokim}}, \bibinfo
  {author} {\bibfnamefont {S.}~\bibnamefont {Rotter}}, \bibinfo {author}
  {\bibfnamefont {R.}~\bibnamefont {El-Ganainy}}, \ and\ \bibinfo {author}
  {\bibfnamefont {{\c S}.~K.}\ \bibnamefont {{\"O}zdemir}},\ }\bibfield
  {title} {\enquote {\bibinfo {title} {Chiral and degenerate perfect absorption
  on exceptional surfaces},}\ }\href@noop {} {\bibfield  {journal} {\bibinfo
  {journal} {Nat. Commun.}\ }\textbf {\bibinfo {volume} {13}},\ \bibinfo
  {pages} {599} (\bibinfo {year} {2022})}\BibitemShut {NoStop}%
\bibitem [{\citenamefont {Suwunnarat}\ \emph {et~al.}(2022)\citenamefont
  {Suwunnarat}, \citenamefont {Tang}, \citenamefont {Reisner}, \citenamefont
  {Mortessagne}, \citenamefont {Kuhl},\ and\ \citenamefont
  {Kottos}}]{Suwunnarat22CP}%
  \BibitemOpen
  \bibfield  {author} {\bibinfo {author} {\bibfnamefont {S.}~\bibnamefont
  {Suwunnarat}}, \bibinfo {author} {\bibfnamefont {Y.}~\bibnamefont {Tang}},
  \bibinfo {author} {\bibfnamefont {M.}~\bibnamefont {Reisner}}, \bibinfo
  {author} {\bibfnamefont {F.}~\bibnamefont {Mortessagne}}, \bibinfo {author}
  {\bibfnamefont {U.}~\bibnamefont {Kuhl}}, \ and\ \bibinfo {author}
  {\bibfnamefont {T.}~\bibnamefont {Kottos}},\ }\bibfield  {title} {\enquote
  {\bibinfo {title} {Non-linear coherent perfect absorption in the proximity of
  exceptional points},}\ }\href@noop {} {\bibfield  {journal} {\bibinfo
  {journal} {Commun. Phys.}\ }\textbf {\bibinfo {volume} {5}},\ \bibinfo
  {pages} {5} (\bibinfo {year} {2022})}\BibitemShut {NoStop}%
\bibitem [{\citenamefont {Shabahang}\ \emph {et~al.}(2017)\citenamefont
  {Shabahang}, \citenamefont {Kondakci}, \citenamefont {Villinger},
  \citenamefont {Perlstein}, \citenamefont {{El H}alawany},\ and\ \citenamefont
  {Abouraddy}}]{Shabahang17SR}%
  \BibitemOpen
  \bibfield  {author} {\bibinfo {author} {\bibfnamefont {S.}~\bibnamefont
  {Shabahang}}, \bibinfo {author} {\bibfnamefont {H.~E.}\ \bibnamefont
  {Kondakci}}, \bibinfo {author} {\bibfnamefont {M.~L.}\ \bibnamefont
  {Villinger}}, \bibinfo {author} {\bibfnamefont {J.~D.}\ \bibnamefont
  {Perlstein}}, \bibinfo {author} {\bibfnamefont {A.}~\bibnamefont {{El
  H}alawany}}, \ and\ \bibinfo {author} {\bibfnamefont {A.~F.}\ \bibnamefont
  {Abouraddy}},\ }\bibfield  {title} {\enquote {\bibinfo {title} {Omni-resonant
  optical micro-cavity},}\ }\href@noop {} {\bibfield  {journal} {\bibinfo
  {journal} {Sci. Rep.}\ }\textbf {\bibinfo {volume} {7}},\ \bibinfo {pages}
  {10336} (\bibinfo {year} {2017})}\BibitemShut {NoStop}%
\bibitem [{\citenamefont {Shabahang}\ \emph {et~al.}(2019)\citenamefont
  {Shabahang}, \citenamefont {Jahromi}, \citenamefont {Shiri}, \citenamefont
  {Schepler},\ and\ \citenamefont {Abouraddy}}]{Shabahang19OL}%
  \BibitemOpen
  \bibfield  {author} {\bibinfo {author} {\bibfnamefont {S.}~\bibnamefont
  {Shabahang}}, \bibinfo {author} {\bibfnamefont {A.~K.}\ \bibnamefont
  {Jahromi}}, \bibinfo {author} {\bibfnamefont {A.}~\bibnamefont {Shiri}},
  \bibinfo {author} {\bibfnamefont {K.~L.}\ \bibnamefont {Schepler}}, \ and\
  \bibinfo {author} {\bibfnamefont {A.~F.}\ \bibnamefont {Abouraddy}},\
  }\bibfield  {title} {\enquote {\bibinfo {title} {Toggling between active and
  passive imaging with an omni-resonant micro-cavity},}\ }\href@noop {}
  {\bibfield  {journal} {\bibinfo  {journal} {Opt. Lett.}\ }\textbf {\bibinfo
  {volume} {44}},\ \bibinfo {pages} {1532--1535} (\bibinfo {year}
  {2019})}\BibitemShut {NoStop}%
\bibitem [{\citenamefont {Shiri}\ \emph
  {et~al.}(2020{\natexlab{a}})\citenamefont {Shiri}, \citenamefont {Yessenov},
  \citenamefont {Aravindakshan},\ and\ \citenamefont {Abouraddy}}]{Shiri20OL}%
  \BibitemOpen
  \bibfield  {author} {\bibinfo {author} {\bibfnamefont {A.}~\bibnamefont
  {Shiri}}, \bibinfo {author} {\bibfnamefont {M.}~\bibnamefont {Yessenov}},
  \bibinfo {author} {\bibfnamefont {R.}~\bibnamefont {Aravindakshan}}, \ and\
  \bibinfo {author} {\bibfnamefont {A.~F.}\ \bibnamefont {Abouraddy}},\
  }\bibfield  {title} {\enquote {\bibinfo {title} {Omni-resonant space-time
  wave packets},}\ }\href@noop {} {\bibfield  {journal} {\bibinfo  {journal}
  {Opt. Lett.}\ }\textbf {\bibinfo {volume} {45}},\ \bibinfo {pages}
  {1774--1777} (\bibinfo {year} {2020}{\natexlab{a}})}\BibitemShut {NoStop}%
\bibitem [{\citenamefont {Shiri}\ \emph
  {et~al.}(2020{\natexlab{b}})\citenamefont {Shiri}, \citenamefont {Schepler},\
  and\ \citenamefont {Abouraddy}}]{Shiri20APLP}%
  \BibitemOpen
  \bibfield  {author} {\bibinfo {author} {\bibfnamefont {A.}~\bibnamefont
  {Shiri}}, \bibinfo {author} {\bibfnamefont {K.~L.}\ \bibnamefont {Schepler}},
  \ and\ \bibinfo {author} {\bibfnamefont {A.~F.}\ \bibnamefont {Abouraddy}},\
  }\bibfield  {title} {\enquote {\bibinfo {title} {Programmable omni-resonance
  using space-time fields},}\ }\href@noop {} {\bibfield  {journal} {\bibinfo
  {journal} {APL Photon.}\ }\textbf {\bibinfo {volume} {5}},\ \bibinfo {pages}
  {106107} (\bibinfo {year} {2020}{\natexlab{b}})}\BibitemShut {NoStop}%
\bibitem [{\citenamefont {Villinger}\ \emph {et~al.}(2021)\citenamefont
  {Villinger}, \citenamefont {Shiri}, \citenamefont {Shabahang}, \citenamefont
  {Jahromi}, \citenamefont {Nasr}, \citenamefont {Villinger},\ and\
  \citenamefont {Abouraddy}}]{Villinger21AOM}%
  \BibitemOpen
  \bibfield  {author} {\bibinfo {author} {\bibfnamefont {M.~L.}\ \bibnamefont
  {Villinger}}, \bibinfo {author} {\bibfnamefont {A.}~\bibnamefont {Shiri}},
  \bibinfo {author} {\bibfnamefont {S}~\bibnamefont {Shabahang}}, \bibinfo
  {author} {\bibfnamefont {A.~K.}\ \bibnamefont {Jahromi}}, \bibinfo {author}
  {\bibfnamefont {M.~B.}\ \bibnamefont {Nasr}}, \bibinfo {author}
  {\bibfnamefont {C.}~\bibnamefont {Villinger}}, \ and\ \bibinfo {author}
  {\bibfnamefont {A.~F.}\ \bibnamefont {Abouraddy}},\ }\bibfield  {title}
  {\enquote {\bibinfo {title} {Doubling the near-infrared photocurrent in a
  solar cell via omni-resonant coherent perfect absorption},}\ }\href@noop {}
  {\bibfield  {journal} {\bibinfo  {journal} {Adv. Opt. Mat.}\ }\textbf
  {\bibinfo {volume} {9}},\ \bibinfo {pages} {2001107} (\bibinfo {year}
  {2021})}\BibitemShut {NoStop}%
\bibitem [{\citenamefont {Jahromi}\ \emph {et~al.}(2021)\citenamefont
  {Jahromi}, \citenamefont {Villinger}, \citenamefont {{El H}alawany},
  \citenamefont {Shabahang}, \citenamefont {Kondakci}, \citenamefont
  {Perlstein},\ and\ \citenamefont {Abouraddy}}]{Jahromi21arxiv}%
  \BibitemOpen
  \bibfield  {author} {\bibinfo {author} {\bibfnamefont {A.~K.}\ \bibnamefont
  {Jahromi}}, \bibinfo {author} {\bibfnamefont {M.~L.}\ \bibnamefont
  {Villinger}}, \bibinfo {author} {\bibfnamefont {A.}~\bibnamefont {{El
  H}alawany}}, \bibinfo {author} {\bibfnamefont {S.}~\bibnamefont {Shabahang}},
  \bibinfo {author} {\bibfnamefont {H.~E.}\ \bibnamefont {Kondakci}}, \bibinfo
  {author} {\bibfnamefont {J.~D.}\ \bibnamefont {Perlstein}}, \ and\ \bibinfo
  {author} {\bibfnamefont {A.~F.}\ \bibnamefont {Abouraddy}},\ }\bibfield
  {title} {\enquote {\bibinfo {title} {Broadband omni-resonant coherent perfect
  absorption in graphene},}\ }\href@noop {} {\bibfield  {journal} {\bibinfo
  {journal} {arXiv:2104.08706}\ } (\bibinfo {year} {2021})}\BibitemShut
  {NoStop}%
\bibitem [{\citenamefont {Shabahang}\ \emph {et~al.}(2021)\citenamefont
  {Shabahang}, \citenamefont {Jahromi}, \citenamefont {Pye}, \citenamefont
  {Perlstein}, \citenamefont {Villinger},\ and\ \citenamefont
  {Abouraddy}}]{Shabahang21JO}%
  \BibitemOpen
  \bibfield  {author} {\bibinfo {author} {\bibfnamefont {S.}~\bibnamefont
  {Shabahang}}, \bibinfo {author} {\bibfnamefont {A.~K.}\ \bibnamefont
  {Jahromi}}, \bibinfo {author} {\bibfnamefont {L.~N.}\ \bibnamefont {Pye}},
  \bibinfo {author} {\bibfnamefont {J.~D.}\ \bibnamefont {Perlstein}}, \bibinfo
  {author} {\bibfnamefont {M.~L.}\ \bibnamefont {Villinger}}, \ and\ \bibinfo
  {author} {\bibfnamefont {A.~F.}\ \bibnamefont {Abouraddy}},\ }\bibfield
  {title} {\enquote {\bibinfo {title} {Coherent perfect absorption in resonant
  materials},}\ }\href@noop {} {\bibfield  {journal} {\bibinfo  {journal} {J.
  Opt.}\ }\textbf {\bibinfo {volume} {23}},\ \bibinfo {pages} {035401}
  (\bibinfo {year} {2021})}\BibitemShut {NoStop}%
\bibitem [{\citenamefont {Arnaud}(1969)}]{Arnaud69AO}%
  \BibitemOpen
  \bibfield  {author} {\bibinfo {author} {\bibfnamefont {J.~A.}\ \bibnamefont
  {Arnaud}},\ }\bibfield  {title} {\enquote {\bibinfo {title} {Degenerate
  optical cavities},}\ }\href@noop {} {\bibfield  {journal} {\bibinfo
  {journal} {Appl. Opt.}\ }\textbf {\bibinfo {volume} {8}},\ \bibinfo {pages}
  {189--196} (\bibinfo {year} {1969})}\BibitemShut {NoStop}%
\bibitem [{\citenamefont {Tradonsky}\ \emph {et~al.}(2019)\citenamefont
  {Tradonsky}, \citenamefont {Gershenzon}, \citenamefont {Pal}, \citenamefont
  {Chriki}, \citenamefont {Friesem}, \citenamefont {Raz},\ and\ \citenamefont
  {Davidson}}]{Tradonsky19SA}%
  \BibitemOpen
  \bibfield  {author} {\bibinfo {author} {\bibfnamefont {C.}~\bibnamefont
  {Tradonsky}}, \bibinfo {author} {\bibfnamefont {I.}~\bibnamefont
  {Gershenzon}}, \bibinfo {author} {\bibfnamefont {V.}~\bibnamefont {Pal}},
  \bibinfo {author} {\bibfnamefont {R.}~\bibnamefont {Chriki}}, \bibinfo
  {author} {\bibfnamefont {A.~A.}\ \bibnamefont {Friesem}}, \bibinfo {author}
  {\bibfnamefont {O.}~\bibnamefont {Raz}}, \ and\ \bibinfo {author}
  {\bibfnamefont {N.}~\bibnamefont {Davidson}},\ }\bibfield  {title} {\enquote
  {\bibinfo {title} {Rapid laser solver for the phase retrieval problem},}\
  }\href@noop {} {\bibfield  {journal} {\bibinfo  {journal} {Sci. Adv.}\
  }\textbf {\bibinfo {volume} {5}},\ \bibinfo {pages} {eaax4530} (\bibinfo
  {year} {2019})}\BibitemShut {NoStop}%
\bibitem [{\citenamefont {Torres}\ \emph {et~al.}(2010)\citenamefont {Torres},
  \citenamefont {Hendrych},\ and\ \citenamefont {Valencia}}]{Torres10AOP}%
  \BibitemOpen
  \bibfield  {author} {\bibinfo {author} {\bibfnamefont {J.~P.}\ \bibnamefont
  {Torres}}, \bibinfo {author} {\bibfnamefont {M.}~\bibnamefont {Hendrych}}, \
  and\ \bibinfo {author} {\bibfnamefont {A.}~\bibnamefont {Valencia}},\
  }\bibfield  {title} {\enquote {\bibinfo {title} {Angular dispersion: an
  enabling tool in nonlinear and quantum optics},}\ }\href@noop {} {\bibfield
  {journal} {\bibinfo  {journal} {Adv. Opt. Photon.}\ }\textbf {\bibinfo
  {volume} {2}},\ \bibinfo {pages} {319--369} (\bibinfo {year}
  {2010})}\BibitemShut {NoStop}%
\bibitem [{\citenamefont {Arbabi}\ \emph {et~al.}(2017)\citenamefont {Arbabi},
  \citenamefont {Arbabi}, \citenamefont {Kamali}, \citenamefont {Horie},\ and\
  \citenamefont {Faraon}}]{Arbabi17Optica}%
  \BibitemOpen
  \bibfield  {author} {\bibinfo {author} {\bibfnamefont {E.}~\bibnamefont
  {Arbabi}}, \bibinfo {author} {\bibfnamefont {A.}~\bibnamefont {Arbabi}},
  \bibinfo {author} {\bibfnamefont {S.~M.}\ \bibnamefont {Kamali}}, \bibinfo
  {author} {\bibfnamefont {Y.}~\bibnamefont {Horie}}, \ and\ \bibinfo {author}
  {\bibfnamefont {A.}~\bibnamefont {Faraon}},\ }\bibfield  {title} {\enquote
  {\bibinfo {title} {Controlling the sign of chromatic dispersion in
  diffractive optics with dielectric metasurfaces},}\ }\href@noop {} {\bibfield
   {journal} {\bibinfo  {journal} {Optica}\ }\textbf {\bibinfo {volume} {4}},\
  \bibinfo {pages} {625--632} (\bibinfo {year} {2017})}\BibitemShut {NoStop}%
\bibitem [{\citenamefont {Geary}(2002)}]{GearyBook2002}%
  \BibitemOpen
  \bibfield  {author} {\bibinfo {author} {\bibfnamefont {J.~M.}\ \bibnamefont
  {Geary}},\ }\href@noop {} {\emph {\bibinfo {title} {Introduction to Lens
  Design}}}\ (\bibinfo  {publisher} {Willmann-Bell},\ \bibinfo {year}
  {2002})\BibitemShut {NoStop}%
\bibitem [{\citenamefont {Shiri}\ and\ \citenamefont
  {Abouraddy}(2022)}]{Shiri22OL}%
  \BibitemOpen
  \bibfield  {author} {\bibinfo {author} {\bibfnamefont {A.}~\bibnamefont
  {Shiri}}\ and\ \bibinfo {author} {\bibfnamefont {A.~F.}\ \bibnamefont
  {Abouraddy}},\ }\bibfield  {title} {\enquote {\bibinfo {title} {Spatial
  resolution of omni-resonant imaging},}\ }\href@noop {} {\bibfield  {journal}
  {\bibinfo  {journal} {Opt. Lett.}\ }\textbf {\bibinfo {volume} {47}},\
  \bibinfo {pages} {3804--3807} (\bibinfo {year} {2022})}\BibitemShut {NoStop}%
\bibitem [{\citenamefont {Hall}\ and\ \citenamefont
  {Abouraddy}(2021)}]{Hall21PRA}%
  \BibitemOpen
  \bibfield  {author} {\bibinfo {author} {\bibfnamefont {L.~A.}\ \bibnamefont
  {Hall}}\ and\ \bibinfo {author} {\bibfnamefont {A.~F.}\ \bibnamefont
  {Abouraddy}},\ }\bibfield  {title} {\enquote {\bibinfo {title} {Spectrally
  recycling space-time wave packets},}\ }\href@noop {} {\bibfield  {journal}
  {\bibinfo  {journal} {Phys. Rev. A}\ }\textbf {\bibinfo {volume} {103}},\
  \bibinfo {pages} {023517} (\bibinfo {year} {2021})}\BibitemShut {NoStop}%
\bibitem [{\citenamefont {Villinger}\ \emph {et~al.}(2015)\citenamefont
  {Villinger}, \citenamefont {Bayat}, \citenamefont {Pye},\ and\ \citenamefont
  {Abouraddy}}]{Villinger15OL}%
  \BibitemOpen
  \bibfield  {author} {\bibinfo {author} {\bibfnamefont {M.~L.}\ \bibnamefont
  {Villinger}}, \bibinfo {author} {\bibfnamefont {M.}~\bibnamefont {Bayat}},
  \bibinfo {author} {\bibfnamefont {L.~N.}\ \bibnamefont {Pye}}, \ and\
  \bibinfo {author} {\bibfnamefont {A.~F.}\ \bibnamefont {Abouraddy}},\
  }\bibfield  {title} {\enquote {\bibinfo {title} {Analytical model for
  coherent perfect absorption in one-dimensional photonic structures},}\
  }\href@noop {} {\bibfield  {journal} {\bibinfo  {journal} {Opt. Lett.}\
  }\textbf {\bibinfo {volume} {40}},\ \bibinfo {pages} {5550--5553} (\bibinfo
  {year} {2015})}\BibitemShut {NoStop}%
\bibitem [{\citenamefont {Pye}\ \emph {et~al.}(2017)\citenamefont {Pye},
  \citenamefont {Villinger}, \citenamefont {Shabahang}, \citenamefont {Larson},
  \citenamefont {Martin},\ and\ \citenamefont {Abouraddy}}]{Pye17OL}%
  \BibitemOpen
  \bibfield  {author} {\bibinfo {author} {\bibfnamefont {L.~N.}\ \bibnamefont
  {Pye}}, \bibinfo {author} {\bibfnamefont {M.~L.}\ \bibnamefont {Villinger}},
  \bibinfo {author} {\bibfnamefont {S.}~\bibnamefont {Shabahang}}, \bibinfo
  {author} {\bibfnamefont {W.~D.}\ \bibnamefont {Larson}}, \bibinfo {author}
  {\bibfnamefont {L.}~\bibnamefont {Martin}}, \ and\ \bibinfo {author}
  {\bibfnamefont {A.~F.}\ \bibnamefont {Abouraddy}},\ }\bibfield  {title}
  {\enquote {\bibinfo {title} {Octave-spanning coherent perfect absorption in a
  thin silicon film},}\ }\href@noop {} {\bibfield  {journal} {\bibinfo
  {journal} {Opt. Lett.}\ }\textbf {\bibinfo {volume} {42}} (\bibinfo {year}
  {2017})}\BibitemShut {NoStop}%
\bibitem [{\citenamefont {Makri}\ \emph {et~al.}(2014)\citenamefont {Makri},
  \citenamefont {Ramezani}, \citenamefont {Kottos},\ and\ \citenamefont
  {Vitebskiy}}]{Makri14PRA}%
  \BibitemOpen
  \bibfield  {author} {\bibinfo {author} {\bibfnamefont {E.}~\bibnamefont
  {Makri}}, \bibinfo {author} {\bibfnamefont {H.}~\bibnamefont {Ramezani}},
  \bibinfo {author} {\bibfnamefont {T.}~\bibnamefont {Kottos}}, \ and\ \bibinfo
  {author} {\bibfnamefont {I.}~\bibnamefont {Vitebskiy}},\ }\bibfield  {title}
  {\enquote {\bibinfo {title} {Concept of a reflective power limiter based on
  nonlinear localized modes},}\ }\href@noop {} {\bibfield  {journal} {\bibinfo
  {journal} {Phys. Rev. A}\ }\textbf {\bibinfo {volume} {89}},\ \bibinfo
  {pages} {031802(R)} (\bibinfo {year} {2014})}\BibitemShut {NoStop}%
\bibitem [{\citenamefont {Vella}\ \emph {et~al.}(2016)\citenamefont {Vella},
  \citenamefont {Goldsmith}, \citenamefont {Browning}, \citenamefont
  {Limberopoulos}, \citenamefont {Vitebskiy}, \citenamefont {Makri},\ and\
  \citenamefont {Kottos}}]{Vella16PRAppl}%
  \BibitemOpen
  \bibfield  {author} {\bibinfo {author} {\bibfnamefont {J.~H.}\ \bibnamefont
  {Vella}}, \bibinfo {author} {\bibfnamefont {J.~H.}\ \bibnamefont
  {Goldsmith}}, \bibinfo {author} {\bibfnamefont {A.~T.}\ \bibnamefont
  {Browning}}, \bibinfo {author} {\bibfnamefont {N.~I.}\ \bibnamefont
  {Limberopoulos}}, \bibinfo {author} {\bibfnamefont {I.}~\bibnamefont
  {Vitebskiy}}, \bibinfo {author} {\bibfnamefont {E.}~\bibnamefont {Makri}}, \
  and\ \bibinfo {author} {\bibfnamefont {T.}~\bibnamefont {Kottos}},\
  }\bibfield  {title} {\enquote {\bibinfo {title} {Experimental realization of
  a reflective optical limiter},}\ }\href@noop {} {\bibfield  {journal}
  {\bibinfo  {journal} {Phys. Rev. Appl.}\ }\textbf {\bibinfo {volume} {5}},\
  \bibinfo {pages} {064010} (\bibinfo {year} {2016})}\BibitemShut {NoStop}%
\bibitem [{\citenamefont {Fang}\ \emph {et~al.}(2015)\citenamefont {Fang},
  \citenamefont {MacDonald},\ and\ \citenamefont {Zheludev}}]{Fang15Light}%
  \BibitemOpen
  \bibfield  {author} {\bibinfo {author} {\bibfnamefont {X.}~\bibnamefont
  {Fang}}, \bibinfo {author} {\bibfnamefont {K.~F.}\ \bibnamefont {MacDonald}},
  \ and\ \bibinfo {author} {\bibfnamefont {N.~I.}\ \bibnamefont {Zheludev}},\
  }\bibfield  {title} {\enquote {\bibinfo {title} {Controlling light with light
  using coherent metadevices: all-optical transistor, summator and invertor},}\
  }\href@noop {} {\bibfield  {journal} {\bibinfo  {journal} {Light Sci. \&
  Appl.}\ }\textbf {\bibinfo {volume} {4}},\ \bibinfo {pages} {e292} (\bibinfo
  {year} {2015})}\BibitemShut {NoStop}%
\bibitem [{\citenamefont {Banerji}\ \emph {et~al.}(2019)\citenamefont
  {Banerji}, \citenamefont {Meem}, \citenamefont {Majumder}, \citenamefont
  {Vasquez}, \citenamefont {Sensale-Rodriguez},\ and\ \citenamefont
  {Menon}}]{Banerji19Optica}%
  \BibitemOpen
  \bibfield  {author} {\bibinfo {author} {\bibfnamefont {S.}~\bibnamefont
  {Banerji}}, \bibinfo {author} {\bibfnamefont {M.}~\bibnamefont {Meem}},
  \bibinfo {author} {\bibfnamefont {A.}~\bibnamefont {Majumder}}, \bibinfo
  {author} {\bibfnamefont {F.~G.}\ \bibnamefont {Vasquez}}, \bibinfo {author}
  {\bibfnamefont {B.}~\bibnamefont {Sensale-Rodriguez}}, \ and\ \bibinfo
  {author} {\bibfnamefont {R.}~\bibnamefont {Menon}},\ }\bibfield  {title}
  {\enquote {\bibinfo {title} {Imaging with flat optics: metalenses or
  diffractive lenses?}}\ }\href@noop {} {\bibfield  {journal} {\bibinfo
  {journal} {Optica}\ }\textbf {\bibinfo {volume} {6}},\ \bibinfo {pages}
  {805--810} (\bibinfo {year} {2019})}\BibitemShut {NoStop}%
\bibitem [{\citenamefont {Meem}\ \emph {et~al.}(2020)\citenamefont {Meem},
  \citenamefont {Banerji}, \citenamefont {Pies}, \citenamefont {Oberbiermann},
  \citenamefont {Majumder}, \citenamefont {Sensale-Rodriguez},\ and\
  \citenamefont {Menon}}]{Meem20Optica}%
  \BibitemOpen
  \bibfield  {author} {\bibinfo {author} {\bibfnamefont {M.}~\bibnamefont
  {Meem}}, \bibinfo {author} {\bibfnamefont {S.}~\bibnamefont {Banerji}},
  \bibinfo {author} {\bibfnamefont {C.}~\bibnamefont {Pies}}, \bibinfo {author}
  {\bibfnamefont {T.}~\bibnamefont {Oberbiermann}}, \bibinfo {author}
  {\bibfnamefont {A.}~\bibnamefont {Majumder}}, \bibinfo {author}
  {\bibfnamefont {B.}~\bibnamefont {Sensale-Rodriguez}}, \ and\ \bibinfo
  {author} {\bibfnamefont {R.}~\bibnamefont {Menon}},\ }\bibfield  {title}
  {\enquote {\bibinfo {title} {Large-area, high-numerical-aperture multi-level
  diffractive lens via inverse design},}\ }\href@noop {} {\bibfield  {journal}
  {\bibinfo  {journal} {Optica}\ }\textbf {\bibinfo {volume} {7}},\ \bibinfo
  {pages} {252--253} (\bibinfo {year} {2020})}\BibitemShut {NoStop}%
\bibitem [{\citenamefont {Nikolov}\ \emph {et~al.}(2021)\citenamefont
  {Nikolov}, \citenamefont {Bauer}, \citenamefont {Cheng}, \citenamefont
  {Kato}, \citenamefont {Vamivakas},\ and\ \citenamefont
  {Rolland}}]{Nikolov21SA}%
  \BibitemOpen
  \bibfield  {author} {\bibinfo {author} {\bibfnamefont {D.~K.}\ \bibnamefont
  {Nikolov}}, \bibinfo {author} {\bibfnamefont {A.}~\bibnamefont {Bauer}},
  \bibinfo {author} {\bibfnamefont {F.}~\bibnamefont {Cheng}}, \bibinfo
  {author} {\bibfnamefont {H.}~\bibnamefont {Kato}}, \bibinfo {author}
  {\bibfnamefont {A.~N.}\ \bibnamefont {Vamivakas}}, \ and\ \bibinfo {author}
  {\bibfnamefont {J.~P.}\ \bibnamefont {Rolland}},\ }\bibfield  {title}
  {\enquote {\bibinfo {title} {Metaform optics: {B}ridging nanophotonics and
  freeform optics},}\ }\href@noop {} {\bibfield  {journal} {\bibinfo  {journal}
  {Sci. Adv.}\ }\textbf {\bibinfo {volume} {7}},\ \bibinfo {pages} {eabe5112}
  (\bibinfo {year} {2021})}\BibitemShut {NoStop}%
\bibitem [{\citenamefont {Yu}\ and\ \citenamefont {Capasso}(2014)}]{Yu14NM}%
  \BibitemOpen
  \bibfield  {author} {\bibinfo {author} {\bibfnamefont {N.}~\bibnamefont
  {Yu}}\ and\ \bibinfo {author} {\bibfnamefont {F.}~\bibnamefont {Capasso}},\
  }\bibfield  {title} {\enquote {\bibinfo {title} {Flat optics with designer
  metasurfaces},}\ }\href@noop {} {\bibfield  {journal} {\bibinfo  {journal}
  {Nat. Mater.}\ }\textbf {\bibinfo {volume} {13}},\ \bibinfo {pages}
  {139--150} (\bibinfo {year} {2014})}\BibitemShut {NoStop}%
\bibitem [{\citenamefont {Dorrah}\ and\ \citenamefont
  {Capasso}(2022)}]{Dorrah22Science}%
  \BibitemOpen
  \bibfield  {author} {\bibinfo {author} {\bibfnamefont {A.~H.}\ \bibnamefont
  {Dorrah}}\ and\ \bibinfo {author} {\bibfnamefont {F.}~\bibnamefont
  {Capasso}},\ }\bibfield  {title} {\enquote {\bibinfo {title} {Tunable
  structured light with flat optics},}\ }\href@noop {} {\bibfield  {journal}
  {\bibinfo  {journal} {Science}\ }\textbf {\bibinfo {volume} {376}},\ \bibinfo
  {pages} {eabi6860} (\bibinfo {year} {2022})}\BibitemShut {NoStop}%
\bibitem [{\citenamefont {Miller}(2023)}]{Miller23Science}%
  \BibitemOpen
  \bibfield  {author} {\bibinfo {author} {\bibfnamefont {D.~A.~B.}\
  \bibnamefont {Miller}},\ }\bibfield  {title} {\enquote {\bibinfo {title} {Why
  optics needs thickness},}\ }\href@noop {} {\bibfield  {journal} {\bibinfo
  {journal} {Science}\ }\textbf {\bibinfo {volume} {379}},\ \bibinfo {pages}
  {41--45} (\bibinfo {year} {2023})}\BibitemShut {NoStop}%
\end{thebibliography}%

\clearpage

\section*{Methods}

\subsection*{Conventional FP resonance at oblique incidence}

The spectral transmission of a collimated optical field from free space at a wavelength $\lambda$ incident on a symmetric FP cavity at an angle $\varphi$ with respect to the cavity normal is:
\begin{equation}
T(\lambda,\varphi)=\frac{1}{\{1+\left(\frac{2\mathcal{F}}{\pi}\right)^{2}\sin^{2}\frac{\chi}{2}},
\end{equation}
where $\mathcal{F}\!=\!\tfrac{\pi\sqrt{R}}{1-R}$ is the cavity finesse, $R$ the cavity mirror reflectivity, the roundtrip cavity phase is:
\begin{equation}
\chi(\lambda,\varphi)=\frac{4\pi d}{\lambda}\sqrt{n^{2}-\sin^{2}\varphi},
\end{equation}
$d$ is the cavity length, and $n$ is its refractive index; we have ignored the phases incurred upon reflection from the cavity mirrors. Resonance occurs at normal incidence at wavelengths $\lambda_{m}\!=\!\tfrac{2nd}{m}$, which correspond to a round-trip phase $\chi(\lambda_{m},0)\!=\!2\pi m$, where integer $m$ is the resonance order [Fig.~\ref{Fig:Theory}(a)]. At oblique incidence, the resonance condition $\chi(\lambda,\varphi)\!=\!2\pi m$ is maintained for the same-order resonance at a blue-shifted wavelength [Fig.~\ref{Fig:Theory}(b,c)]:
\begin{equation}
\lambda_{m}(\varphi)=\lambda_{m}\sqrt{1-\tfrac{1}{n^{2}}\sin^{2}\varphi}<\lambda_{m}.
\end{equation}
Therefore, wavelengths associated with the $m^{\mathrm{th}}$ resonance range from $\lambda\!=\!\lambda_{m}$ at normal incidence to $\lambda\!=\!\lambda_{\mathrm{min}}\!=\!\lambda_{m}(1-\sigma_{n})$ at glancing angle ($\varphi\!\rightarrow\!90^{\circ}$), where we define for convenience the parameter
\begin{equation}
\sigma_{n}=1-\sqrt{1-\tfrac{1}{n^{2}}},
\end{equation}
which depends solely on the cavity refractive index. 

The first- and second-order derivatives of $\varphi(\lambda)$ for the $m^{\mathrm{th}}$-resonance are:
\begin{eqnarray}
\frac{d\varphi}{d\lambda}&=&-\frac{\lambda}{\sqrt{\lambda_{m}^{2}-\lambda^{2}}\sqrt{\lambda^{2}-\lambda_{\mathrm{min}}^{2}}},\label{eq:FirstOrderDerivative}\\\frac{d^{2}\varphi}{d\lambda^{2}}&=&\frac{\lambda_{m}^{2}\lambda_{\mathrm{min}}^{2}-\lambda^{4}}{(\lambda_{m}^{2}-\lambda^{2})^{3/2}(\lambda^{2}-\lambda_{\mathrm{min}}^{2})^{3/2}}.\label{eq:SecondOrderDerivative}
\end{eqnarray}
Both expressions are valid in the spectral range $\lambda_{\mathrm{min}}\!<\!\lambda\!<\!\lambda_{m}$. Setting $\tfrac{d^{2}\varphi}{d\lambda^{2}}\!=\!0$ in Eq.~\ref{eq:SecondOrderDerivative} yields the wavelength:
\begin{equation}
\lambda_{\mathrm{c}}=\sqrt{\lambda_{m}\lambda_{\mathrm{min}}},
\end{equation}
which is the geometric mean of the two extreme wavelengths associated with the $m^{\mathrm{th}}$-resonance. Substituting $\lambda\!=\!\lambda_{\mathrm{c}}$ into Eq.~\ref{eq:FirstOrderDerivative}, we obtain the linear AD coefficient:
\begin{equation}
\beta_{m}=\tfrac{d\varphi}{d\lambda}\big|_{\lambda_{\mathrm{c}}}=-\frac{1}{\lambda_{m}-\lambda_{m}^{\mathrm{min}}}=-\frac{1}{\Delta\lambda_{m}}=-\frac{1}{\lambda_{m}\sigma_{n}}.
\end{equation}
For $m\!=\!11$, the cavity parameters in the main text yield $\Delta\lambda_{m}\!\approx\!158$~nm and $\beta_{11}\!\approx\!-0.36^{\circ}$/nm.

\subsection*{Experimental arrangements for exciting achromatic resonances}

In both setups in Fig.~\ref{Fig:Setups}(a) and Fig.~\ref{Fig:Setups}(d), we make use of a collimated white-light source (Thorlabs MCWHL8), and all the lens apertures are 50~mm in diameter. In the first setup [Fig.~\ref{Fig:Setups}(a)] we use a diffraction grating $G_{1}'$ with groove density 830~lines/mm, area $25\!\times\!25$~mm$^{2}$, an incident angle $20^{\circ}$ with respect to the grating normal, and the -1~diffraction order is selected at an angle of $5^{\circ}$. This configuration provides AD of $\beta\approx-0.05^{\circ}$/nm, which is significantly lower in magnitude than the target value $\beta_{11}\approx-0.36^{\circ}$/nm associated with the $m\!=\!11$ resonance. We increase the value of $\beta$ provided by the grating by a factor $7\times$ before incidence on the FP cavity using an afocal magnification system comprising an achromatic biconvex lens (L$_{1}$; $f\!=\!300$~mm), followed by a plano-convex aspheric condenser lens (L$_{2}'$; $f\!=\!40$~mm; Thorlabs ACL5040U). The distance between G$_{1}'$ and L$_{1}$ is 300~mm, L$_1$ and L$_{2}'$ are separated by 330~mm, and the distance from L$_{2}'$ to the FP cavity is 25~mm. The system is then mirrored after the FP cavity to recombine the wavelengths. We describe below in more detail the AD spectral profile produced by the aspheric lens L$_{2}'$.

In the setup shown in Fig.~\ref{Fig:Setups}(d), light impinges on a diffraction grating G$_{1}$ (1400~lines/mm, $25\!\times\!25$-mm$^{2}$ area) at an angle $20^{\circ}$ with respect to its normal, and the -1 diffracted order is selected (diffracted angle $-25^{\circ}$). The AD produced by the grating in this configuration is $\beta\approx0.09^{\circ}$/nm, which again is insufficient to satisfy the omni-resonance condition. To increase the value of $\beta$, we make use of an afocal imaging telescope configuration whose $4\times$ magnification factor increases the AD in turn by $4\times$ \cite{Shabahang17SR}. To minimize spherical and chromatic aberrations, which is a critical requirement to achieve high-quality color-imaging, we employed a three-lens system. The sequence of lenses comprises an achromat biconvex L$_{1}$ ($f\!=\!300$~mm) followed by two achromat biconvex lenses L$_{2}$ and L$_{3}$ ($f\!=\!100$~mm) in an afocal configuration, with separation 317~mm between L$_{1}$ and L$_{2}$, and 56~mm between L$_{2}$ and L$_{3}$. The FP cavity is placed at 14~mm from L$_{3}$.

\subsection*{Calculating the aberrations of the omni-resonant imaging systems}

Zemax (OpticStudios~22.3) was used to optimize the parameters of the omni-resonant system. We first consider the setup in Fig.~\ref{Fig:Setups}(a) containing an aspheric lens. We optimized the focal lengths of the lenses and the separation between them to reduce the spherical and chromatic aberrations across the omni-resonant bandwidth. The best result associated with the system parameters described above is a spherical aberration of $\sim\!60$~wavelengths with an entrance pupil of 20~mm [Fig.~\ref{Fig:Setups}(b), inset]. The calculated AD profile from this omni-resonant system [plotted in red in Fig.~\ref{Fig:Setups}(b)], and is compared to the spectral trajectory of the conventional FP resonance $m\!=\!11$. In addition, we also plot the curves corresponding to a drop in spectral transmission to $\tfrac{1}{e^{2}}$ of the on-resonance maximum value assuming a finesses of $\mathcal{F}\!=\!10$. We also optimized the focal lengths of the lenses and the distances between them for the setup in Fig.~\ref{Fig:Setups}(d) using Zemax. Here we reach calculated spherical aberrations of $\sim\!0.2$~wavelengths for an entrance pupil of diameter 20~mm, thus producing an Airy radius of 36~$\upmu$m that contains within it the majority of the spectrum [Fig.~\ref{Fig:Setups}(e), inset].

\subsection*{Nonlinear AD provided by the aspheric lens}

We outline here the calculation of the full AD spectral profile $\varphi(\lambda)$ produced by the aspheric lens L$_{2}'$ in the setup in Fig.~\ref{Fig:Setups}(a). This AD profile combines linear and nonlinear contributions [Fig.~\ref{Fig:Setups}(b)], which together increase the omni-resonant bandwidth of the achromatic resonance $m\!=\!11$ [Fig.~\ref{Fig:Setups}(c)]. Referring to Fig.~\ref{Fig:Setups}(b), bottom left inset, the aspheric lens L$_{2}'$ is plano-convex, with light incident on the convex side and emerging from the planar side.

We consider collimated light incident on the aspheric lens, with all the rays parallel to the $z$-axis. We aim at finding the propagation angle $\varphi$ with respect to the $z$-axis after the lens for any incident ray on the lens. The ray coinciding with the $z$-axis (the symmetry axis of the lens) emerges along the $z$-axis. A ray parallel to the $z$-axis but displaced from it vertically by a height $h$ undergoes refraction at both lens surfaces. At the planar surface, refraction follows simply from Snell's law. At the convex surface, refraction depends on the local curvature $R(h)$ at height $h$. If $R$ is a constant, $R(h)\!=\!R_{\mathrm{o}}$, then the lens corresponds to a traditional spherical lens. However, for free-form optics or aspheric lenses, $R$ can vary locally with height. Defining the lens `sag', the height-dependent thickness $z(h)$ of the lens, the local curvature is given by:
\begin{equation}
R(h)=h\sqrt{1+\left(\frac{dz}{dh}\right)^{2}}.
\end{equation}
The specific sag profile for the aspheric lens L$_{2}'$ is given by:
\begin{equation}
z(h)=\frac{h^{2}}{R_{\mathrm{o}}\left(1+\sqrt{1-(1+k)\frac{h^2}{R_{\mathrm{o}}^{2}}}\right)}+A_{4}h^4,
\end{equation}
where $k$ is the conic section, $R_{\mathrm{o}}$ is the local radius curvature at $h\!=\!0$, and $A_4$ is the fourth order coefficient \cite{GearyBook2002}; here we have $R_{\mathrm{o}}\!=\!20$~mm, $k\!=\!-0.6$, and $A_{4}\!=\!2\times10^{-6}$~mm$^{-3}$.

Using this sag profile $z(h)$, we calculate the local curvature $R(h)$, and then evaluate the height-dependent angle emerging from the lens obtained by ray-tracing:
\begin{equation}\label{eq:HeightDependentAngle}
\sin\{\varphi(h)\}=-\frac{h}{R}\sqrt{1-\left(\frac{h}{R}\right)^2}+\frac{h}{R}\sqrt{n_{\mathrm{L}}^{2}-\left(\frac{h}{R}\right)^2},
\end{equation}
where $n_{\mathrm{L}}$ is the refractive index of L$_{2}'$. Note that $\varphi(-h)\!=\!-\varphi(h)$ as required by the symmetry of the lens configuration.

The collimated wave front produced by the lens L$_{1}$ and incident on L$_{2}'$ is spectrally resolved in space. Each ray at height $h$ corresponds to a different wavelength $\lambda$ as dictated by the AD introduced by G$_{1}'$ and the focal length of L$_{1}$:
\begin{equation}\label{eq:GratingAD}
h(\lambda)=\frac{mf}{\Lambda}(\lambda-\lambda_{\mathrm{c}}),
\end{equation}
where $\lambda_{\mathrm{c}}$ is the central wavelength located on-axis at the lens $h\!=\!0$, $m$ is the diffraction order selected after the grating, $f$ is the focal length of L$_{1}$, and $\Lambda$ is the groove density of G$_{1}'$. The angle $\varphi(h)$ in Eq.~\ref{eq:HeightDependentAngle} now becomes the sought-after wavelength-dependent angle $\varphi(\lambda)$ representing the AD profile produced by the aspheric lens L$_{2}'$ and incident on the FP cavity. This AD profile $\varphi(\lambda)$ is plotted in Fig.~\ref{Fig:Setups}(b) and contains both linear and nonlinear terms. In the vicinity of the lens center (wavelengths close to $\lambda_{\mathrm{c}}$), the AD profile is dominated by the linear term, which matches the AD profile of the spectral trajectory for the FP resonance close to $\lambda_{\mathrm{c}}$. As $h$ increases away from the lens center and approaches the edge of the lens L$_{2}'$ (wavelengths far from $\lambda_{\mathrm{c}}$), the angle $\varphi(\lambda)$ deviates from the linear form obtained in the vicinity of $\lambda_{\mathrm{c}}$, and the contribution from nonlinear terms becomes significant. 

\subsection*{Determining the omni-resonant bandwidth}

To determine the omni-resonant bandwidth theoretically, we first evaluate the extent of the spectral width of the spectral trajectory for the selected conventional FP resonance ($m\!=\!11$ here). The exact spectral trajectory of this FP resonance is given by:
\begin{equation}
\sin\{\varphi(\lambda)\}=n\sqrt{1-\left(\frac{\lambda}{\lambda_{m}}\right)^{2}},
\end{equation}
where $\lambda_{m}\!=\!\tfrac{2nd}{m}$. This spectral trajectory corresponds to the peak of the spectral transmission for this symmetric FP cavity, $T(\lambda,\varphi(\lambda))\!=\!1$. We next find the spectral trajectories $\varphi_{\pm}(\lambda)$, corresponding to the reduced transmission condition $T(\lambda,\varphi_{\pm}(\lambda))\!=\!\tfrac{1}{\eta}$:
\begin{equation}
\sin\varphi_{\pm}(\lambda)=n\sqrt{1-\left(\frac{\lambda}{\lambda_{m}^{\pm}}\right)^{2}},
\end{equation}
where the wavelengths $\lambda_{m}^{\pm}$ are given by:
\begin{equation}
\lambda_{m}^{\pm}=\frac{\lambda_{m}}{1\mp\frac{1}{m\pi}\sin^{-1}\left(\frac{\pi}{2\mathcal{F}}\sqrt{\eta-1}\right)}.
\end{equation}
For large finesse $\mathcal{F}\!\gg\!1$, $\lambda_{m}^{\pm}$ can be simplified to:
\begin{equation}
\lambda_{m}^{\pm}\approx\lambda_{m}(1\pm\frac{1}{2m\mathcal{F}}\sqrt{\eta-1}).
\end{equation}
In our case, $\eta\!=\!e^{2}$, $m\!=\!11$, and $\lambda_{m}\!\approx\!70$~nm. In both Fig.~\ref{Fig:Setups}(b) and Fig.~\ref{Fig:Setups}(e) we plot $\varphi_{+}(\lambda)$ and $\varphi_{-}(\lambda)$ as dotted curves around the exact spectral trajectory $\varphi(\lambda)$ for the FP resonance.

To evaluate the omni-resonant bandwidth, we find the wavelengths at which the curves $\varphi_{+}(\lambda)$ and $\varphi_{-}(\lambda)$ intersect with the AD profile $\varphi(\lambda)$ produced by the omni-resonant system. In the case of the setup in Fig.~\ref{Fig:Setups}(d), the AD provides by the system is $\varphi(\lambda)\!=\!\psi+\beta_{11}(\lambda-\lambda_{\mathrm{c}})$. This trajectory $\varphi(\lambda)$ intersects with $\varphi_{+}(\lambda)$ at the wavelength $\lambda_{\mathrm{c}}\!<\!\lambda_{+}\!<\!\lambda_{m}$, and intersects with $\lambda_{-}(\lambda)$ at the wavelength $\lambda_{\mathrm{min}}\!<\!\lambda_{-}\!<\!\lambda_{\mathrm{c}}$. The omni-resonant bandwidth is then estimated to be $\Delta\lambda\!\approx\!\lambda_{+}-\lambda_{-}$. In the case of the setup in Fig.~\ref{Fig:Setups}(a), the AD provided by the omni-resonant system combines both linear and nonlinear contributions, and $\varphi(\lambda)$ here is obtained by combining Eq.~\ref{eq:HeightDependentAngle} with Eq.~\ref{eq:GratingAD}. We then obtain the wavelengths $\lambda_{+}$ and $\lambda_{-}$ at which $\varphi_{+}(\lambda)$ and $\varphi_{-}(\lambda)$ intersect with $\varphi(\lambda)$, respectively, and find the omni-resonant bandwidth $\Delta\lambda\!=\!\lambda_{+}-\lambda_{-}$.

\textbf{Acknowledgments.} This work was funded by the U.S. Office of Naval Research under contracts N00014-17-1-2458 and N00014-20-1-2789.

\subsection*{Supplementary Movies}

\begin{figure*}[t!]
\centering
\includegraphics[width=16cm]{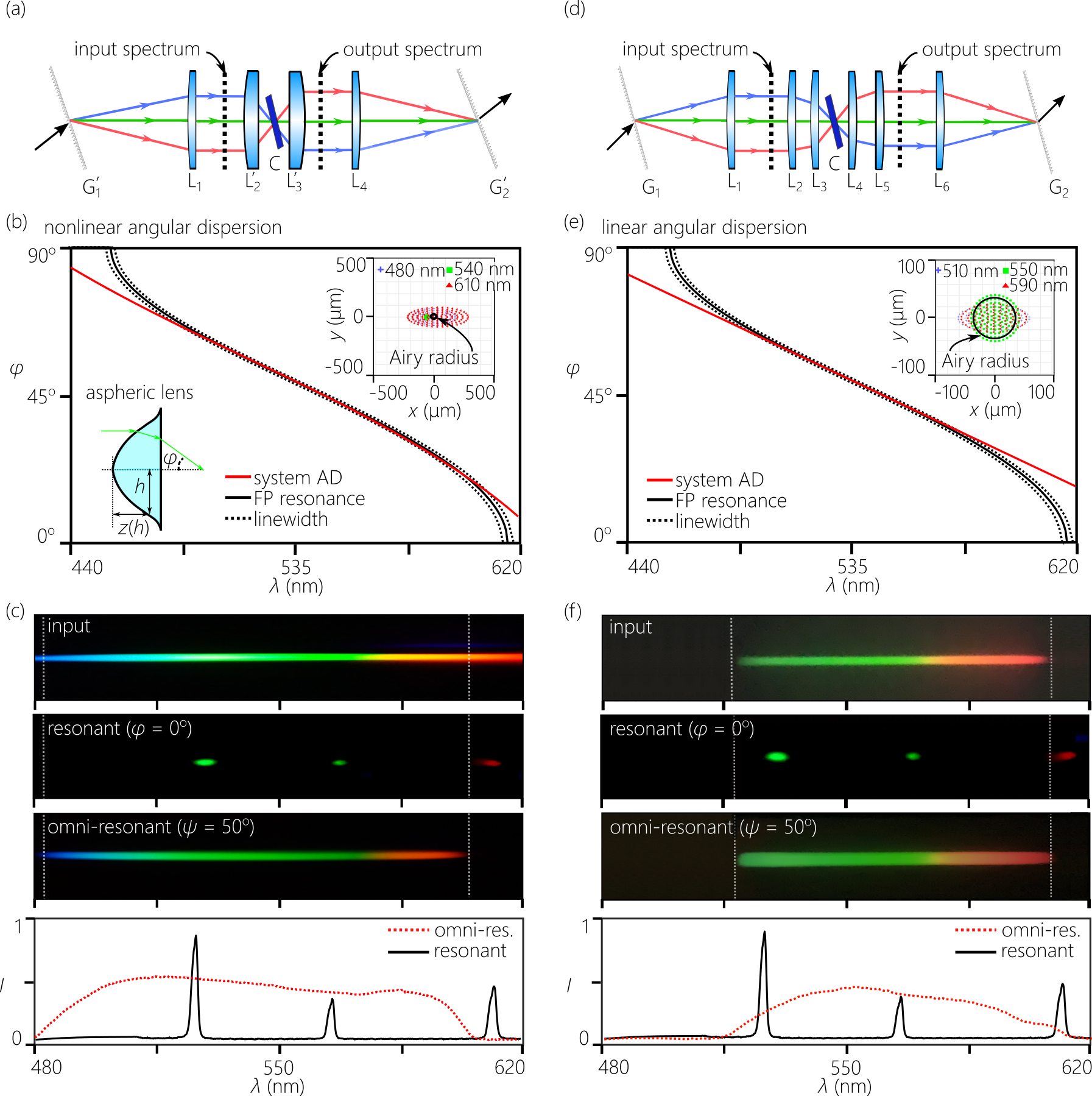}
\caption{\textbf{Omni-resonant system.} (a) Schematic of a broadband omni-resonant system. (b) A comparison between the spectral trajectory of the FP-resonance $m\!=\!11$ (solid black curve), and the AD (solid, red curve) produced by the system in (a). We also show the spectrally broadened trajectory resulting from finite cavity finesse (dotted curves, $\mathcal{F}\!=\!11$ here). Top right inset displays the calculated Airy disk at three wavelengths, revealing significant aberrations beyond the ideal Airy radius. Bottom left inset depicts the geometry of the aspheric lens L$_{2}'$. (c) The measured spectrum at the input plane, the measured resonant transmission of the FP cavity at normal incidence, and the measured omni-resonant spectrum at the output plane. The bottom panel overlays the resonant and omni-resonant spectra. (d-f) Same as (a-c) but utilizing a different configuration to minimize the spherical and chromatic abberations. G: Diffraction grating; L: lens; C: FP cavity.}
\label{Fig:Setups}
\end{figure*}

\begin{figure*}[t!]
\centering
\includegraphics[width=17.6cm]{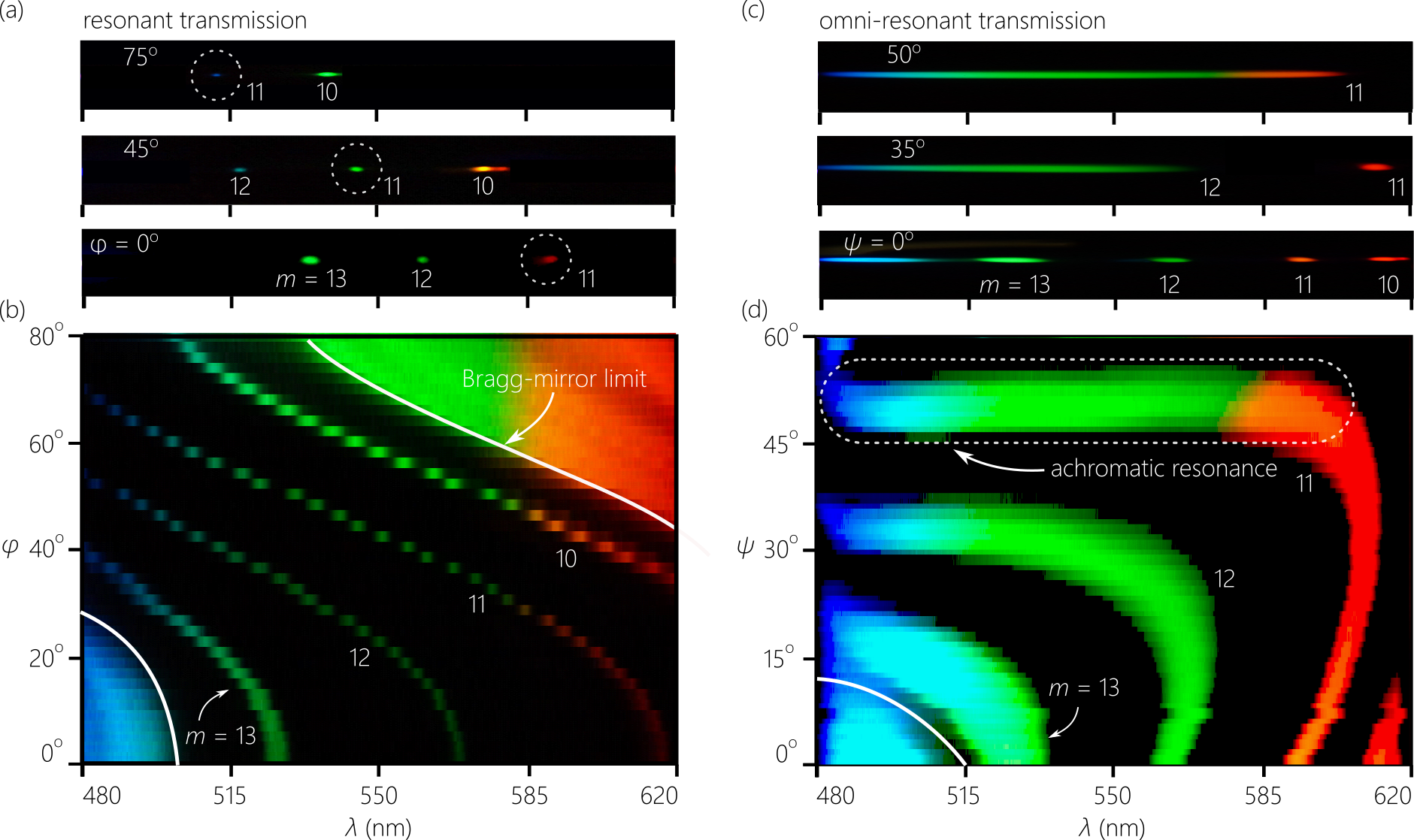}
\caption{\textbf{Measured resonant and omni-resonant spectra.} (a) Measured resonant transmission through the FP cavity at select incident angles $\varphi\!=\!0^{\circ}$, $50^{\circ}$, and $75^{\circ}$. (b) Measured resonant transmission $T(\lambda,\varphi)$ while varying $\varphi$ from $0$ to $80^{\circ}$ (see Supplementary Movie~1). (c) Measured spectral transmission through the FP cavity at the output plane in the setup in Fig.~\ref{Fig:Setups}(a), obtained at select cavity tilt angles $\psi\!=\!0^{\circ}$, $35^{\circ}$, and $50^{\circ}$. (d) Measured omni-resonant transmission $T(\lambda,\psi)$ while varying $\psi$ from $0$ to $60^{\circ}$ (see Supplementary Movie~2 and Supplementary Movie~3). The solid white curves in (b) and (d) correspond to the limits of the reflection spectrum of the Bragg mirrors in the FP cavity.}
\label{Fig:Spectra}
\end{figure*}

\begin{figure*}[t!]
\centering
\includegraphics[width=8.6cm]{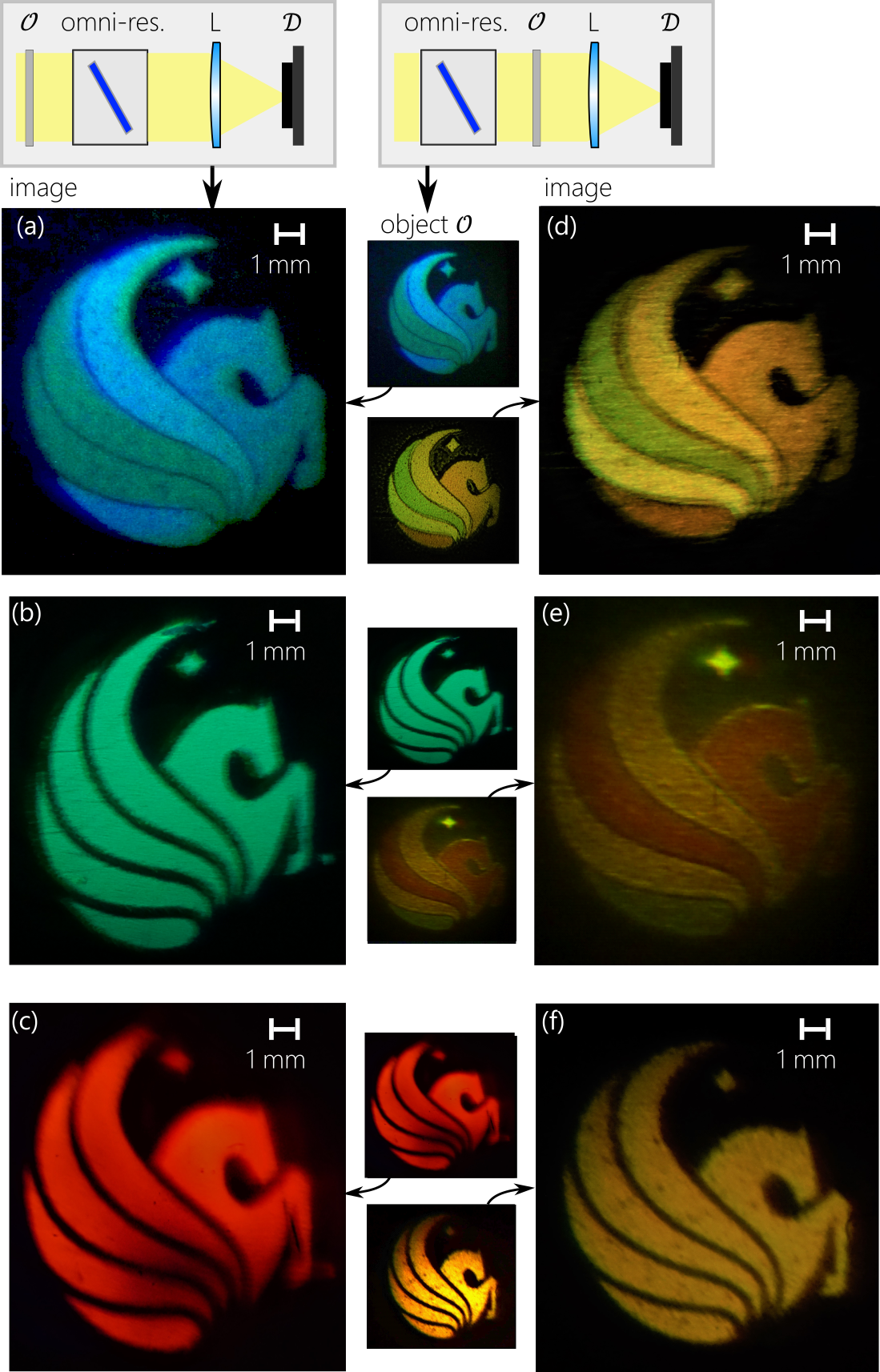}
\caption{\textbf{Broadband omni-resonant imaging in the visible.} (a) Omni-resonant image of an object in the form of a Pegasus that is (a) predominantly blue, (b) green, (c) red, (d,e) multi-colored, and (f) black-and-white. The inset image associated with each omni-resonant image is a reference obtained by illuminating the object with a 100-nm-bandwidth spectrum filtered from the white-light source. The two optical configurations to obtain the omni-resonant and reference images are shown on top. $\mathcal{O}$: Object; L: imaging lens; and $\mathcal{D}$: detection plane. The images in (b-f) make use of the $m=11$ achromatic resonance at $\psi\!\approx\!50^{\circ}$, while (a) makes use of $m=12$ at $\psi\approx35^{\circ}$ [Fig.~\ref{Fig:Spectra}(d)].}
\label{Fig:Pegasus}
\end{figure*}

\textbf{\textit{Supplementary Movie~1.}} Real-time measurement of the conventional spectral response for the FP cavity corresponding to the data plotted in Fig.~\ref{Fig:Spectra}(a). The cavity is continuously tilted by an angle $\varphi$ from $0^{\circ}$ to $80^{\circ}$ with respect to a collimated field whose spectrum is spatially resolved using a combination of the diffraction grating G$_{1}'$ and lens L$_{1}$, corresponding to those in the setup in Fig.~\ref{Fig:Setups}(a). The spectrally resolved field transmitted through the FP cavity is captured by a color camera, so that the wavelength of light is denoted both in the color and in the location along the horizontal axis. At $\varphi\sim0^{\circ}$, the continuous spectrum on the blue side is outside the range of the Bragg mirrors in the FP cavity; similarly for the continuous spectrum on the red side appearing after $\varphi\sim50^{\circ}$.

\textbf{\textit{Supplementary Movie~2.}} Real-time measurement of the omni-resonant spectral response taken at the output plane in the setup in Fig.~\ref{Fig:Setups}(a). The cavity is continuously tilted an angle $\psi$ varying from $0^{\circ}$ to $70^{\circ}$ with respect to the incident field in which AD matching that if the FP-resonance $m=11$ has been introduced. At $\psi\approx50^{\circ}$ the broadest achromatic resonance appears with a bandwidth $\Delta\lambda\approx130$~nm (extending from the edge of the blue to the red). Once again, the transmitted field is captured using a color camera, so that the wavelength is reflected both in the color and the position along the horizontal axis. 

\textbf{\textit{Supplementary Movie~3.}} Same as Supplementary Movie~2 except that the omni-resonant spectrum is captured at the output plane in Fig.~\ref{Fig:Spectra}(d). At $\psi\approx50^{\circ}$ the broadest achromatic resonance appears with a bandwidth $\Delta\lambda\approx100$~nm (extending from the green to the red). 

\begin{figure*}[t!]
\centering
\includegraphics[width=8.6cm]{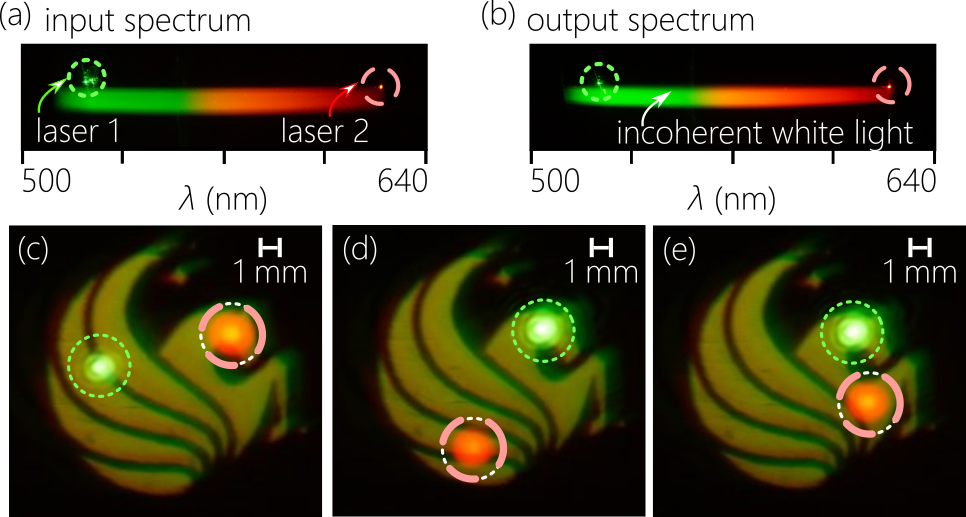}
\caption{\textbf{Omni-resonant imaging combining coherent and incoherent light.} (a) Measured input and (b) output spectra of a combined extended white-light source and localized green (laser~1) and red (laser~2) lasers. (c-e) Omni-resonant images of an object (a black-and-white Pegasus) illuminated with the extended, stationary white-light source and two moving localized green and red lasers (the latter two identified by dashed circles); see Supplementary Movie~4.}
\label{Fig:Movies}
\end{figure*}

\textbf{\textit{Supplementary Movie~4.}} Real-time measurement of omni-resonant image obtained using Fig.~\ref{Fig:Setups}(d) placed in a single-lens imaging system. The object is a black-and-white Pegasus illuminated with spectrally incoherent white-light, and two spatially localized lasers (green and red) are scanned simultaneously but independently across the object plane. Stills from this Movie are displayed in Fig.~\ref{Fig:Movies}(c-e). The omni-resonant image captured here using a color camera shows the stationary object and the moving laser beams in real time.

\clearpage

\clearpage

\clearpage

\clearpage

\clearpage

\clearpage

\clearpage

\end{document}